\documentclass[12pt,english]{article}
\usepackage{amsmath}
\usepackage{graphicx}
\usepackage{epsfig}

\providecommand{\tabularnewline}{\\}

\usepackage{textcomp}

\makeatother

\usepackage{babel}

\begin{document}

\begin{center}

{\bf{\large Determining  the Octant of $\theta_{23}$ at  LBNE in conjunction with Reactor Experiments}}

\vspace{0.4cm}

Kalpana Bora$\mbox{}^{a)}$
\footnote{kalpana.bora@gmail.com},
Debajyoti Dutta$\mbox{}^{a)}$
\footnote{debajyotidutta1985@gmail.com},
Pomita Ghoshal$\mbox{}^{b)}$
\footnote{pomita.ghoshal@gmail.com}

\vspace{0.2cm}

$\mbox{}^{a)}${\em Department of Physics,
Gauhati University, Guwahati- 781 014,\\
Assam, India}

\vspace{0.1cm}
$\mbox{}^{b)}${\em Department of Physics,
LNM Institute of Information Technology (LNMIIT),\\
Rupa-ki-Nangal, post-Sumel, via-Jamdoli,
Jaipur-302 031,
Rajasthan}

\end{center}

\begin{abstract}

We have explored the possibility of resolving the $\theta_{23}$ octant degeneracy present in the newly planned Long-Baseline Neutrino Experiment (LBNE) by combining reactor experiments. Reactor experiments have already measured the third mixing angle $\theta_{13}$ very precisely and this precise measurement of $\theta_{13}$ in turn helps in determining the octant of $\theta_{23}$. We have examined the octant sensitivity with and without data from reactors. Octant sensitivity increases if reactors are added. The comparative study of octant sensitivities of 10 kt and 35 kt liquid argon far detector(FD), with and without the near detector(ND), reveals that the sensitivity increases with an increase in detector mass. Also, the presence of the ND improves the sensitivity. The effect of adding prior on octant determination is also studied in this work.  

\end{abstract}

\section{Introduction}

The precise measurement of $\theta_{13}$ by the three new generation reactor experiments \cite{1,2,3} has opened up a new phase in neutrino physics. The flux of neutrinos from nuclear reactors is very high, and it can be used for precision measurements of unknown parameters in the Standard Model which can lead to new discoveries. The initial result of Daya Bay (DB) experiment confirmed the non-zero value of $\theta_{13}$ with more than 5$\sigma$ precision \cite{4}. Subsequently, RENO also confirmed the non-zero value of $\theta_{13}$ and the best fit value of $\theta_{13}$ at 4.9$\sigma$ \cite{5} is given as $$\sin^22\theta_{13}=0.100\pm0.010(stat)\pm0.015(syst)$$   
The establishment of the fact that $\theta_{13}$ $\sim9^0$ brings us closer to answering many remaining  questions like CP violation in the neutrino as well as leptonic sectors, absolute neutrino masses and mixing, determination of neutrino mass hierarchy, the octant of $\theta_{23}$ etc.\\
After reactor experiments, it is now time for long baseline experiments to hunt for new physics. LBNEs have many advantages due to their long baselines. Earth matter effects, which become important for any terrestrial baseline of several hundred kilometers, have opposite signs in the probability expression for the two hierarchies. Thus, in long-baseline experiments (LBNE \cite{6, 7}, T2K \cite{8}, MINOS \cite{9}, LBNO \cite{10} etc.), due to matter effects, it is possible to differentiate between normal mass hierarchy (NH, $m^2_{3}-m^2_{2}>0$) and inverted mass hierarchy (IH, $m^2_{3}-m^2_{2} < 0$). In the light of the MSW effect \cite{11}, the ability of LBNEs to resolve hierarchy increases significantly. On the other hand, the large and precise value of the neutrino mixing angle $\theta_{13}$ helps in studying the octant of $\theta_{23}$.    

The other two mixing angles $\theta_{12}$ ( by solar neutrino \cite{12}, KamLAND experiment \cite{13}) and $\theta_{23}$ ( by SuperKamiokande\cite{14}) are also precisely measured. But MINOS data contradicts the atmospheric neutrino data dominated by SuperKamiokande. MINOS disappearance measurement still indicates non-maximal $\theta_{23}$ while Super-Kamiokande-dominated atmospheric data prefers a maximal value of $\theta_{23}$. T2K disappearance measurements provide the most precise value of atmospheric mixing angle $\theta_{23}$ and it prefers a best fit value which is very close to maximal \cite{15}. But after adding data from reactors, the global best fit value of $\theta_{23}$ moves to the higher octant (HO: $\theta_{23}$ $ >$ 45$^0$) \cite{15} (for lower octant(LO), $\theta_{23}$ $ <$ 45$^0$). Precise measurement of $\theta_{13}$ is responsible for this. The recent global fit points towards the higher octant for both NH and IH hierarchies. 

 Parameter degeneracy is one of the main problems of long baseline experiments. Due to the inherent structure of three flavour neutrino oscillation probabilities, several disconnected regions appear in the multi-dimensional neutrino oscillation parameter space. The appearance of several disconnected regions for a given set of true values makes it difficult to pin-point the exact or true solution. These eight-fold parameter degeneracies \cite{16,17,18, 19} are classified as intrinsic or ($\delta_{cp},\theta_{13}$)-degeneracy \cite{20, 21}, hierarchy or sign ($\Delta{m}^{2}$)-degeneracy \cite{22, 23} and  octant or ($\theta_{23}$) -degeneracy \cite{15, 18, 23}.  
The appearance of solutions corresponding to $\theta_{23}$ lying in the two different octants
at different values of $\delta_{cp}$ and $\theta_{13}$ makes it difficult to point out the true octant. This is known as octant degeneracy. 

In this work, we have studied the octant sensitivity as well as the possibility of resolving the octant degeneracy at the newly planned Long-Baseline Neutrino Experiment (LBNE). We have also studied the effects of the near detector(ND) on octant sensitivity. In the literature, there are many extensive studies regarding the octant ambiguity. How the matter effect in long-baseline experiments can help in resolving octant degeneracy has been studied in ref \cite{24}. Recently, the possibility of resolving octant degeneracy by combining T2K and NOvA \cite{25} experiment has been studied in ref. \cite{26}. In ref. \cite{27}, the impact of adding atmospheric data with T2K and NOvA to improve the octant sensitivity is studied. The possibility of the possibility of observing CP violation in the proposed INO experiment, using the beta
beam from CERN \cite{28}. How the combination of different baselines can help in resolving parameter degeneracies is studied in ref. \cite{29}. The effect of non-zero $\theta_{13}$ in resolving parameter degeneracies in LBL is studied in \cite{30}. In \cite{31}, parameter degeneracies as well as some other ambiguities are studied by combining various reactor experiments with J-PARC(T2K) and NuMI off-axis (NOvA) data. In \cite{32}, the signal for CP violation is searched in reactors as well as in superbeam experiments considering non-zero $\theta_{13}$. The phenomenology of atmospheric neutrinos associated
with the deviation of the 2-3 leptonic mixing from maximal is discussed in \cite{33}. The possibility of resolving octant ambiguity by a direct measurement of the sign of D$\equiv 1- \sin^2_{23}$ utilizing
atmospheric neutrino data is studied in \cite{34}. Neutrino oscillation probabilities as well as the possibility of determining the neutrino oscillation parameters from different experiments is studied in \cite{35}. In \cite{36}, the physics potential of Hyper-Kamiokande Experiment with an underground Water Cherenkov detector is discussed. The possibility of resolving mass hierarchy and octant with atmospheric neutrinos is also studied in \cite{37}.

In this work, we have separately combined each of the three reactor experiments Double 
Chooz (DC), DB and RENO with LBNE to examine the improved sensitivity. We have also checked the possibility of resolving the octant ambiguity in the light of precisely measured $\theta_{13}$ by combining the three reactors with LBNE. Reactor experiments have measured $\theta_{13}$ by using $\bar{\nu}_e$ disappearance channel while LBNE is capable of measuring $\theta_{13}$ with $\bar{\nu}_e$ and $\nu_e$ channels. Without any external constraint on $\theta_{13}$, LBNE can measure $\sin^2 2\theta_{13}$ \cite{5}. We have presented our octant sensitivity results with and without ND for both 10 kt and 35 kt liquid argon detector (LBNE). We also study how the presence of a prior on $\theta_{13}$ affects the octant determination capability of LBNE. The paper has been organized as follows: in Section 2, we have explained how and why the octant degeneracy arises and in Section 3, we have provided the details of relevant experiments for completeness. In Section 4, we have explained the statistical procedure as well as the $\chi^2$ analysis used to produce the results. In Section 5, we have presented the figures and results. Conclusions are presented in Section 6.

\section{Octant degeneracy from Theoretical Point of View} 

As mentioned above, the octant degeneracy is due to the inherent structure of three-flavour neutrino oscillation probability. Long-baseline neutrino experiments are sensitive to $\nu_{\mu}\rightarrow\nu_{\mu}$, $\nu_{\mu}\rightarrow\nu_{e}/\nu_{\tau}$ oscillations. In this section, we analyse the capacity of $\nu_{\mu}$ disappearance and $\nu_{e}$ appearance signals to probe octant ambiguity.

  The series expansion of the survival probability $P(\nu_{\mu}\rightarrow\nu_{\mu})$ in matter upto the second order in $\alpha$ is \cite{38, 39, 40}
  
  $$ P^m_{\mu\mu}=1-\sin^22\theta_{23}\sin^2\Delta +\alpha\, c^2_{12}\,\sin^22\theta_{23}\Delta\sin2\Delta-\alpha^2\sin^22\theta_{12}\,c^2_{23}\frac{\sin^2A\Delta}{A^2}-\alpha^2\,c^4_{12}\sin^22\theta_{23}\Delta^2\cos2\Delta $$
  $$+\frac{1}{2A}\alpha^2\sin^22\theta_{12}\sin^22\theta_{23}\,(\sin\Delta\frac{\sin A\Delta}{A}\cos(A-1)\Delta-\frac{\Delta}{2}\sin2\Delta)$$
  $$-4s^2_{13}\,s^2_{23}\frac{\sin^2(A-1)\Delta}{(A-1)^2}
  -\frac{2}{A-1}\,s^2_{13}\,\sin^22\theta_{23}(\sin\Delta\cos A\Delta\frac{\sin(A-1)\Delta}{A-1}-\frac{A}{2}\Delta\sin2\Delta)$$
  $$-2\alpha\,s_{13}\sin2\theta_{12}\sin2\theta_{23}\cos\delta_{cp}\cos\Delta\frac{\sin A\Delta}{A}\frac{\sin(A-1)\Delta}{A-1}$$
  \begin{equation}
  \label{a}
 +\frac{2}{A-1}\alpha\,s_{13}\sin2\theta_{12}\sin2\theta_{23}\cos2\theta_{23}\cos\delta_{cp}\sin\Delta(A\sin\Delta-
  \frac{\sin A\Delta}{A}\cos(A-1)\Delta)
   \end{equation}
  
 where c and s stand for cosine and sine. We define $$\Delta\equiv\frac{\bigtriangleup m^2_{31}L}{4E},\, \alpha \equiv\frac{\bigtriangleup m^2_{21}}{\bigtriangleup m^2_{31}},\,\, A\equiv\frac{2EV}{\Delta m^2_{31}}.$$ Here $V=\sqrt{2}\,G_F\,n_e$ and A is the MSW matter potential, $G_{F}$ is Fermi coupling constant, $n_{e}$ is the number density of electrons and E is the neutrino energy. In vacuum, this expression for $P(\nu_{\mu}\rightarrow\nu_{\mu})$ survival probability reduces to 
   $$ P^v_{\mu\mu}=1-\sin^22\theta_{23}\sin^2\Delta +\alpha\, c^2_{12}\,\sin^22\theta_{23}\Delta\sin2\Delta-\alpha^2\Delta^2[\sin^22\theta_{12}c^2_{23}+c^2_{12}\sin^22\theta_{23}(\cos2\Delta-s^2_{12})]$$
 \begin{equation}
 \label{b}
 + 4 s^2_{13} s^2_{23}\cos2\theta_{23}\sin^2\Delta-2\alpha \,s_{13} \sin2\theta_{12} \, s^2_{23} \sin2\theta_{23}\cos\delta_{cp}\Delta\sin2\Delta
 \end{equation}
 
  In the one mass square dominant approximation (OMSDA)\cite{38, 41}, the smaller mass squared difference
  $\bigtriangleup m^2_{21}$ is neglected compared to the $\bigtriangleup m^2_{31}$. The OMSDA condition is \begin{equation}
  \frac{\Delta m^2_{21}L}{4E}<<1\,\,\, or\,\,\, \frac{L}{E}<<10^4 km/GeV 
   \end{equation}
  This condition is not valid for small $\theta_{13}$ because the terms with small $\Delta m^2_{21}$ can be dropped only if they are small compared to the leading order terms with $\theta_{13}$. In OMSDA, survival probability in constant matter density can be expressed as
 
$$P^m_{\mu\mu}=1-\cos^2\theta^m_{13}\sin^22\theta_{23}\sin^2[1.27\,\bigtriangleup m^2_{31}(\frac{1+A+
\frac{(\bigtriangleup m^2_{31})^m}{\bigtriangleup m^2_{31}}}{2})\frac{L}{E}]$$
$$-\sin^2\theta^m_{13}\sin^22\theta_{23}\sin^2[1.27\,\bigtriangleup m^2_{31}(\frac{1+A-
\frac{(\bigtriangleup m^2_{31})^m}{\bigtriangleup m^2_{31}}}{2})\frac{L}{E}]$$

\begin{equation}
\label{c}
-\sin^4 \theta_{23}\sin^22\theta^m_{13}\sin^2[1.27(\bigtriangleup m^2_{31})^m)\frac{L}{E}]
\end{equation}
where
$$(\bigtriangleup m^2_{31})^m = \bigtriangleup m^2_{31}[(\cos2\theta_{13}-A)^2+(\sin2\theta_{13})^2]^{\frac{1}{2}}$$
\begin{equation}
\sin2\theta^m_{13} = \frac{\sin2\theta_{13}}{[(\cos2\theta_{13}-A)^2+(\sin2\theta_{13})^2]^{\frac{1}{2}}}
\end{equation}
In vacuum, (\ref{b}) reduces to  

\begin{equation}
\label{d}
P^v_{\mu\mu}=1-\sin^22\theta_{23}\,\sin^2[1.27 \,\Delta m^2_{31}\,\frac{L}{E}] + 4s^2_{13}\,s^2_{23} \cos2\theta_{23}\,\sin^2[1.27 \,\Delta m^2_{31}\,\frac{L}{E}]
\end{equation} 
 
 It is observed from the (\ref{a}), (\ref{b}), (\ref{c}) and (\ref{d}) that the leading terms are dependent on $\sin^22\theta_{23}$. In the case of $P^v_{\mu\mu}$ this leads to intrinsic octant degeneracy i.e. the dominant term in the $\nu_{\mu}$ disappearance measurement is sensitive to $\sin^22\theta_{23}$ and hence for a given value of the survival probability P($\nu_{\mu}\rightarrow\nu_{\mu}$), we have two-fold solutions for $\theta_{23}$. In terms of probability, it can be written as:
\begin{equation}P(\theta_{23}) = P(\pi/2-\theta_{23})
\end{equation}
i.e. when a measurement is made, it cannot differentiate between $\theta_{23}$ and $\pi/2-\theta_{23}$.\\
The series expansion of $P(\nu_e\rightarrow \nu_\mu)$ oscillation probability in matter upto second order in $\alpha$ can be written as  \cite{38}
$$
P^m_{e\mu}=\alpha^2\,\sin^22\theta_{12}\,c^2_{23}\,\frac{\sin^2\,A\Delta}{A^2}+4s^2_{13}\,s^2_{23}\,\frac{\sin^2 (A-1)\Delta}{(A-1)^2}$$
\begin{equation}
\label{e}
+2\alpha\,s_{13}\,\sin2\theta_{12}\,
\sin2\theta_{23}\,\cos(\Delta-\delta_{cp})\frac{\sin A\Delta}{A}\frac{\sin(A-1)\Delta}{A-1}
\end{equation}
In vacuum, (\ref{e}) changes to
\begin{equation}
\label{f}
P^v_{e\mu}=\alpha^2\,\sin^22\theta_{12}\,c^2_{23}\,\Delta^2+4s^2_{13}\,s^2_{23}\,\sin^2\Delta+2\alpha\,s_{13}\,\sin2\theta_{12}\,
\sin2\theta_{23}\,\cos(\Delta-\delta_{cp})\Delta \sin\Delta
\end{equation}
The probability $P^m_{e\mu}$ is same as $P^m_{\mu e}$ with the replacement $\delta_{cp}\rightarrow -\delta_{cp}$. But in OMSDA,  $P^m_{e\mu}=P^m_{\mu e}$ as the probability in this case is insensitive to $\delta_{cp}$.

The appearance of $\nu_e$ in $\nu_\mu$ beam ($P_{\mu e}$) in vacuum and in matter under OMSDA approximation and in the light of precise value of $\theta_{13}$ is given as \cite{38, 41}

\begin{equation}
\label{g}
P^v_{e\mu}=P^v_{\mu e} = \sin^2 \theta_{23}\sin^22\theta_{13}\sin^2[1.27(\bigtriangleup m^2_{31})\frac{L}{E}]
\end{equation}
\begin{equation}
\label{h}
P^m_{e\mu}=P^m_{\mu e} = \sin^2 \theta_{23}\sin^22\theta^m_{13}\sin^2[1.27(\bigtriangleup m^2_{31})^m\frac{L}{E}]
\end{equation}

 It is observed that the leading terms in $P^v_{\mu e}$ and $P^m_{\mu e}$ depend on $\sin^2\theta_{23}$ but still (\ref{e}), (\ref{f}), (\ref{g}) and (\ref{h}) cannot measure $\sin^2\theta_{23}$ independent of $\sin^22\theta_{13}$ for a given $\bigtriangleup m^2_{31}$. Instead, (\ref{g}) measures the quantity $\sin^2 \theta_{23}\sin^22\theta_{13}$ \cite{28}. In matter, the quantity $\sin2\theta^m_{13}$ approaches to 1 and hence the combined quantity $\sin^2 \theta_{23}\sin^22\theta^m_{13}$ is no longer invariant \cite{27}. So the octant degeneracy can be lifted. The higher order corrections of (\ref{g}) and (\ref{h}) depend on $\delta_{cp}$ as shown in (\ref{e}) and (\ref{f}). Determination of octant of $\theta_{23}$ is affected by the octant-$\delta_{cp}$ degeneracy which can be probed with sufficient $\nu$ and $\bar{\nu}$ 
data \cite{26}. The probability function for different values of $\theta_{13}$ and $\delta_{cp}$ can be written as:
\begin{equation}P(\theta_{23},\theta_{13},\delta_{cp}) = P(\theta_{23}', \theta_{13}',\delta_{cp}')\end{equation}
Where, $\theta_{23}'$ lies in the opposite octant of $\theta_{23}$.

Reactor experiments are disappearance experiments and they measure $P(\bar{\nu_{e}}\rightarrow\bar{\nu_{e}})$ survival probability. As the baselines of the reactor experiments are very small, so matter effect is negligible. Under the assumption that $\frac{\bigtriangleup m^2_{21}}{\bigtriangleup m^2_{31}}$ is small, we can approximate the survival probability for reactor experiments as \cite{42, 43}:
\begin{equation}
P(\bar{\nu_{e}}\rightarrow\bar{\nu_{e}})=1-\sin^{2}2\theta_{13}\sin^{2}(\frac{\Delta{m}_{31}^{2}L}{4E})-\cos^{4}\theta_{13}\sin^{2}2\theta_{12}\sin^{2}(\frac{\Delta{m}^2_{21}L}{4E})\end{equation}

Reactor experiments can precisely measure the third mixing angle $\theta_{13}$ for a given value of the mass-squared difference $\bigtriangleup m^2_{31}$. So the newly proposed long-baseline neutrino experiment (LBNE), which is rich in $\nu$ and $\bar{\nu}$ and can measure both $\nu_{\mu}$ disappearance and $\nu_e$ appearance probability, if combined with reactor experiments, can lift the octant degeneracy. 

\section{Details of the experiments}

In this section, we present some technical details regarding LBNE and the three reactor experiments Double Chooz, DB and RENO for the sake of completeness.
 
 LBNE \cite{5, 6} is a US based future accelerator experiment which has been developed to probe many significant questions in the neutrino sector. In the 1st phase of LBNE, which is called LBNE10, a very intense on-axis beam of neutrinos from Fermilab is planned to Homestake mine where a non magnetized 10 kt Liquid Argon (LAr) Time Projection Chamber will detect neutrinos from the beam. The baseline is 1300 km. The ND, employed at Fermilab, will help in controlling the systematic uncertainties. The ND will also measure the absolute flux as well as the energy dependence of all the neutrino species to predict the far/near flux ratio as a function of energy. We have considered 100 kt-yr exposure for the 700 kW beam over a period of 10 years with 5 years in neutrino mode and 5 years in anti-neutrino mode. The presence of the ND reduces the background systematics at FD. The values of the systematics considered with and without ND are listed in Table 1 and Table 2 \cite{6, 44, 45}.

 \begin{table}[!h]
\begin{center}
\begin{tabular}{|l|c|r|l|}
\hline 
Channel & Signal normalisation  & BG normalisation & Calibration   \tabularnewline
  &    error &       error & error  \tabularnewline
\hline 
$\nu_{\mu}(\bar{\nu_{\mu}})\rightarrow\nu_{e}(\bar{\nu_e})$ & 1$\%$  & 1$\%$ & 2$\%$ \tabularnewline
\hline 
$\nu_{\mu}(\bar{\nu_{\mu}})\rightarrow\nu_{\mu}(\bar{\nu_{\mu}})$ & 1$\%$   & 5$\%$  & 2$\%$  \tabularnewline
\hline 
\end{tabular}
\caption{Systematics at FD in presence of ND in LBNE}
\label{Table 2}

\par\end{center}
\end{table}

 \begin{table}[!h]
\begin{center}
\begin{tabular}{|l|c|r|l|}
\hline 
Channel & Signal normalisation  & BG normalisation & Calibration   \tabularnewline
  &    error &       error & error  \tabularnewline
\hline  
$\nu_{\mu}(\bar{\nu_{\mu}})\rightarrow\nu_{e}(\bar{\nu_e})$ & 5$\%$  & 10$\%$ & 2$\%$ \tabularnewline
\hline 
$\nu_{\mu}(\bar{\nu_{\mu}})\rightarrow\nu_{\mu}(\bar{\nu_{\mu}})$ & 5$\%$   & 45$\%$  & 2$\%$  \tabularnewline
\hline 
\end{tabular}
\caption{Systematics at FD without ND in LBNE}
\label{Table 2}

\par\end{center}
\end{table}
 
 The Double Chooz (DC) \cite{3, 46} reactor experiment in France is designed to detect $\bar{\nu}_{e}$
through the reaction-\begin{equation}
\bar{\nu}_{e}+p\rightarrow e^{+}+n\end{equation} known as inverse beta decay reaction.
 The source of anti-neutrino flux is the $\beta$ decay of the fission products of
four main isotopes- $^{235}U$, $^{239}Pu$, $^{241}Pu$ and $^{238}U$ at Chooz
power plant. DC has two reactor cores and the distances of the far and near detectors are listed in Table 3.

\begin{table}[!h]
\begin{center}
\begin{tabular}{|l|c|r|}
\hline 
Reactor No  & ND(km)  & FD(km) \tabularnewline
\hline 
1  & 0.47  & 1.12 \tabularnewline
\hline 
2  & 0.35  & 1.00 \tabularnewline
\hline 
\end{tabular}
\caption{Core- detector distances in DC}
\label{Table 3}

\par\end{center}
\end{table} 
 Both ND and FD are identical and the mass of each detector is about 10.16 tons. The reactor site contains two reactors. The thermal power of each reactor is 4.27 GW.
 
 The Daya Bay (DB) \cite{1,47} neutrino experiment in China has three nuclear power
plants (NPPs): the Daya Bay NPP, the Ling Ao NPP, and the Ling Ao II NPP. The Ling Ao II NPP has been working since 2010-2011. Each site has two reactor cores. During normal operation, each core generates 2.9 GW of power and hence total power output of six cores is 17.4 GW. DB has two ND sites and one FD site. Each Near site has two detectors while the far site has four detectors. The mass of each detector
is about 20 tons. The Daya Bay ND site is located at 363 m from
the center of the Daya Bay cores while Ling Ao near
detector hall is at 481 m from the center of the Ling Ao cores, and 526 m from the center of the Ling Ao II cores. The FD is situated at a distance of 1985 m from the midpoint of the Daya Bay cores. The distance from the mid point of the Ling Ao and Ling Ao II cores to the FD is 1615 m.

The RENO \cite{2,42} experiment at Korea is designed to search for reactor anti-neutrino disappearance using two identical detectors. The RENO set-up has a ND roughly 292 m away and the FD is about 1.4 km away from the reactor
array center. Each detector mass is 16 tons. There are six reactor cores in RENO, for which the core to detector distances are listed in table 4.

\begin{table}[!h]
\begin{center}
\begin{tabular}{|l|c|r|}
\hline 
Reactor No  & FD(km) & ND(km)  \tabularnewline
\hline 
1  &  1.52 & 0.70 \tabularnewline
\hline 
2  & 1.43 & 0.48   \tabularnewline
\hline 
3  &  1.39 & 0.32   \tabularnewline
\hline 
4  &  1.39 & 0.32  \tabularnewline
\hline 
5  & 1.43 & 0.48   \tabularnewline
\hline 
6    & 1.52 & 0.70  \tabularnewline
\hline

\end{tabular}
\caption{Core-detector distances in RENO}
\label{Table 4}

\par\end{center}
\end{table}

In this analysis, we have made a comparative study of the octant sensitivity of the 10 kt and 35 kt FD with and without ND. For all the three reactors, we have considered the same exposure i.e. exposure of 3 years for anti-neutrinos.  We have used a reduced set of systematic
errors as documented in the respective experimental literature \cite{42, 46, 47}. Taking into account the partial cancellation of errors due to the presence of both ND and FD, the values of errors are listed in Table 5.
\begin{table}[!t]
\begin{center}
\begin{tabular}{|l|l|c|c|}
\hline 
Name of Exp & RENO & Double-Chooz  & Daya Bay   \tabularnewline
\hline 
Reactor correlated error($\%$)  & 0.5 & 0.06  & 0.12  \tabularnewline
\hline 
Detector normalisation error($\%$) & 0.5  & 0.06  & 0.12  \tabularnewline
\hline 
Scaling or calibration error($\%$) & 0.1  & 0.5  & 0.5  \tabularnewline
\hline 
Overall normalization error($\%$) & 0.5  & 0.5  & 0.5  \tabularnewline
\hline 
Isotopic abundance error($\%$) & 0.5  & 0.06  & 0.12  \tabularnewline
\hline
\end{tabular}
\caption{Reduced set of errors used in our calculation}
\par\end{center}
\end{table}

\section{Statistical details and $\chi^2$ analysis}

  In this work, we have used GLoBES \cite{48, 49} for simulating the experiments. The total no of energy bins in each of the four experiments is 40. For LBNE, energy range is 1-10 GeV while that for the reactor experiments is 1.8-8 MeV. Event reconstruction efficiency of 85$\%$ is considered in the simulation for LBNE beamline (i.e. for both $\nu_e$ and $\nu_{\mu}$ channels) in neutrino and anti-neutrino mode. All other information like signal efficiency, background efficiency, resolution function etc. for LBNE is taken from \cite{6}. In the simulation of three reactor experiments, we have considered an uncertainty of 2$\%$ associated with the shape of neutrino energy spectrum. Energy resolution used is 12$\%$ for all three reactors. We have checked one sample result with the true resolutions of the reactors (6.5$\%$ for RENO, 8$\%$ for DB and 7.5$\%$ for DC at 1 $\rm MeV$), and we find that the effect of the reactor energy resolution is very negligible. While performing $\chi^2$ analysis for reactor experiments, we have to take care of all the systematic uncertainties associated in the experiments. The $\chi^2$ function we have used here for simulating the reactor experiments is taken from GLoBES manual and is given by:
 \begin{equation}\chi^{2}=\underset{i=1}{\sum^{\#bins}}\underset{d=N,F}{\sum}\frac{(O_{d,i}-(1+a_{R}+a_{d})T_{d,i})^{2}}{O_{d,i}}+\frac{a_{R}^{2}}{\sigma_{R}^{2}}+\frac{a_{N}^{2}}{\sigma_{N}^{2}}+\frac{a_{F}^{2}}{\sigma_{F}^{2}}\end{equation}

where $a_{R}, a_{F},a_{N}$ are uncertainties associated with reactor flux
and detector mass (F stands for FD and N for ND) and $\sigma_{R}, $ $\sigma_{F}, \sigma_{N}$ are respective standard deviations. $O_{N,i}$, $O_{F,i}$ denote the event rates for the i-th bin in ND and FD, calculated for true values of oscillation parameters while T$_{d,i}$ are the expected event rates for the i-th bin in the far and near detector for the test values.

$\chi^2$ in this analysis is calculated between the true octant and the wrong octant for the true values of $\delta_{cp}$. When the true values of $\theta_{23}$ are in the lower octant (higher octant), the test value runs in the higher octant (lower octant) which is the so-called wrong octant.
To examine the octant sensitivity, we have first calculated the minimum $\chi^2$ for LBNE marginalizing over the parameter set. Then in the combined analysis, $\chi^2$ of each of the reactors is added with LBNE and minimum $\chi^2$ is calculated for the combined set. We have considered the best fit values of the oscillation parameters \cite{50} as the true values. The recent best fit values of the oscillation parameters \cite{15} have changed slightly. We have checked one sample set of results with the new best fit values (fig. 2), and it is found that the result is changing slightly with the new data. At $\theta_{23}=36^0$ and $54^0$ the sensitivity increases slightly with the new global data. But both $36^0$ and $54^0$ are outside the $3\sigma$  range of $\theta_{23}$. We have shown these two points just for the consistency of the figures. Interestingly, it is observed that the octant sensitivity slightly decreases with the new global data at $3\sigma$ cl.

\begin{figure}[!h]
\begin{centering}
\begin{tabular}{cc}
\includegraphics[width=0.5\columnwidth]{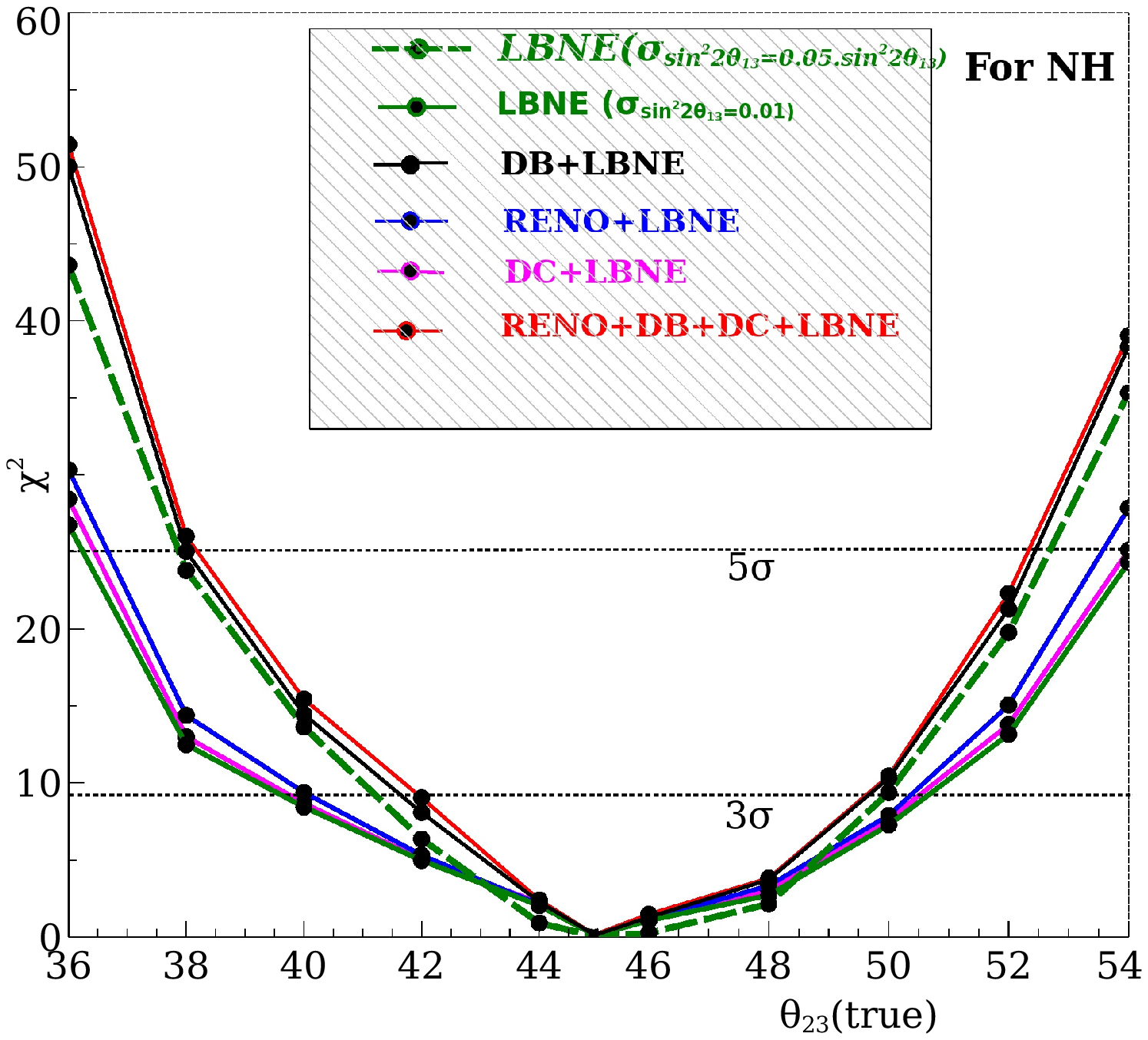}
\includegraphics[width=0.5\columnwidth]{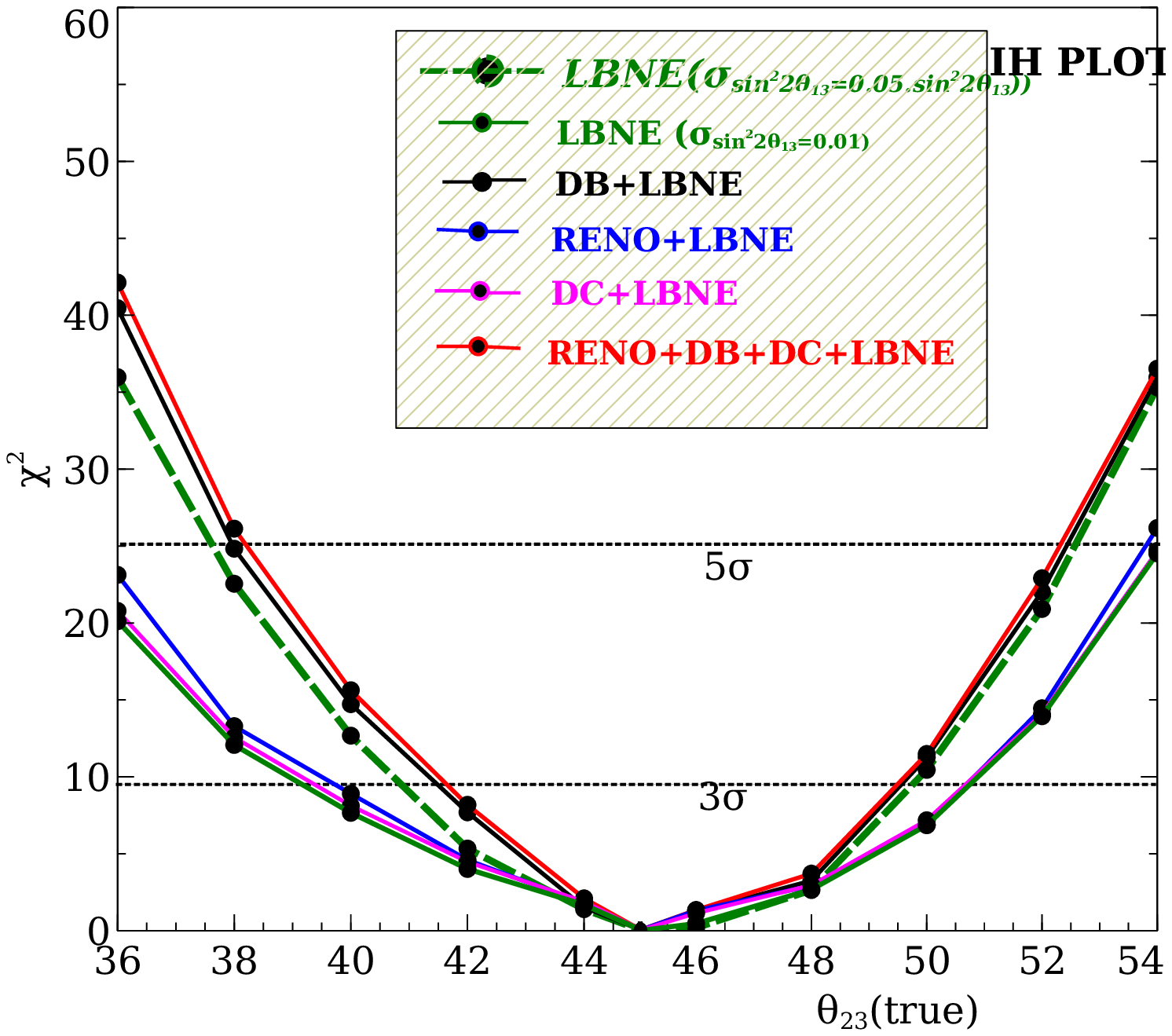}\\
\includegraphics[width=0.5\columnwidth]{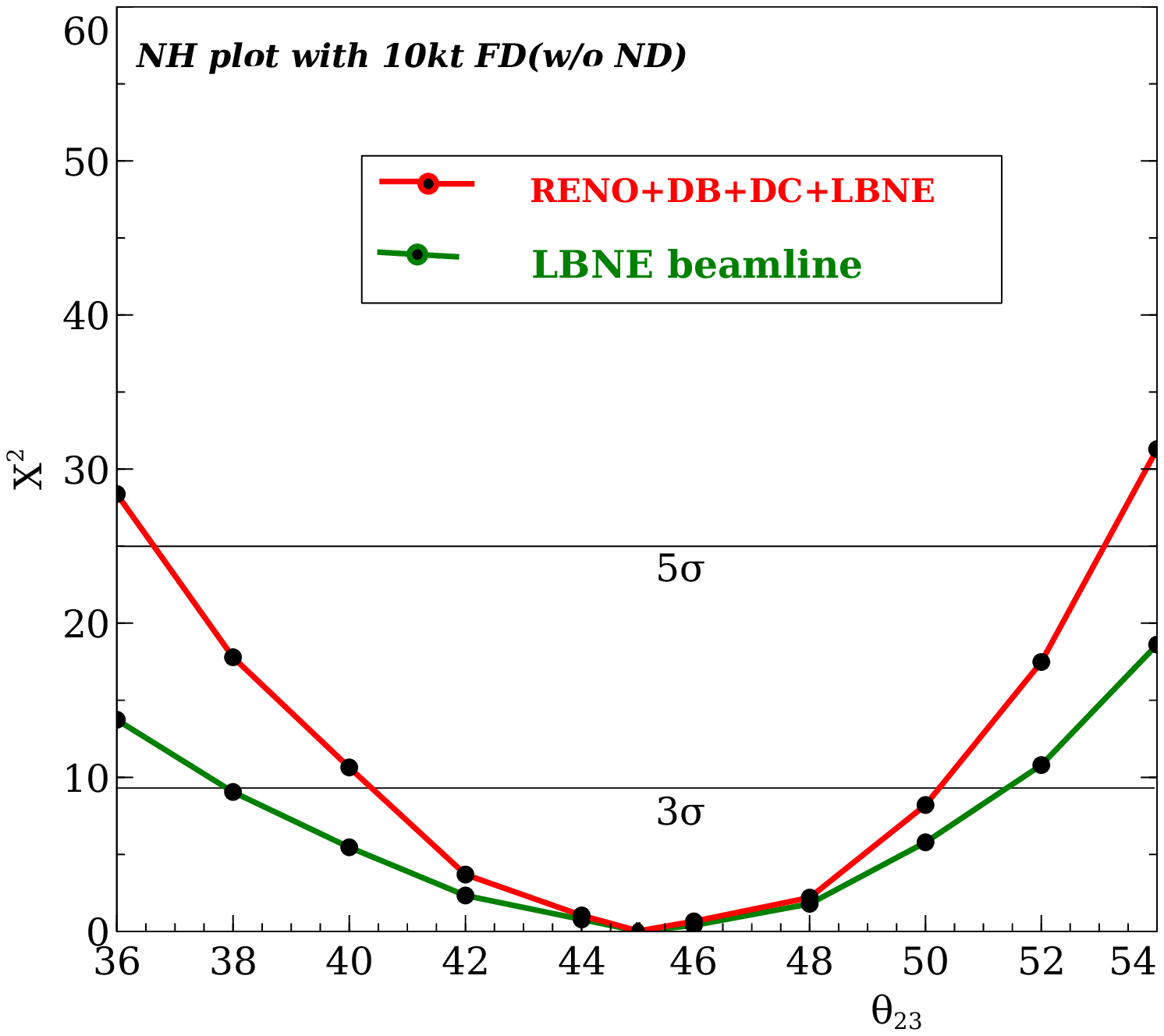}
\includegraphics[width=0.5\columnwidth]{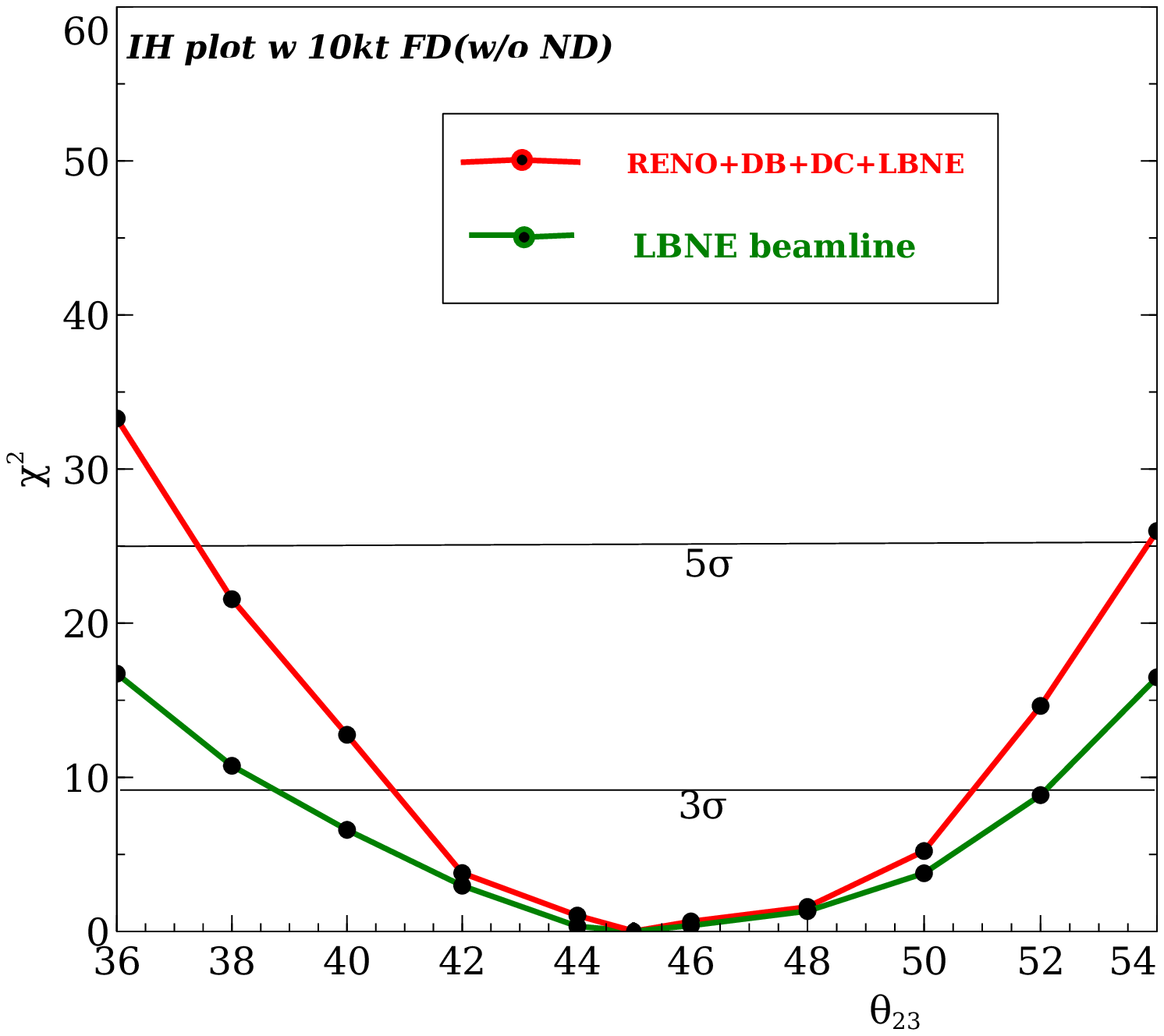}
\end{tabular}
\par\end{centering} 

\caption{ Octant sensitivity plots for LBNE(10 kt FD) with (upper panel) ND and without (lower panel) ND. In the upper panel, we have presented the sensitivity for two different values of added prior on $\theta_{13}$ in presence of ND. The Green dashed plot (upper panel) is with a prior of $\sigma_{\sin^22\theta_{13}} = 0.05 \times\sin^22\theta_{13} $ and the thin green plot is with a prior of $\sigma_{\sin^22\theta_{13}} = 0.01$. In the lower panel, we have presented the results for only one value of added prior on $\theta_{13}$, i.e. $\sigma_{\sin^22\theta_{13}} = 0.01$ without ND. In the upper panel, three reactor experiments DC, RENO and DB are separately added with LBNE and finally we have combined all the three reactors with LBNE to examine the octant sensitivity. In the lower panel, we have compared the octant sensitivity of the combined (RENO+DB+DC+LBNE) experiments with that of LBNE (w/o ND for LBNE). We have combined 3 years of $\bar{\nu}$ data from each reactor experiments with the (5+5) years of LBNE data in $\nu$ and $\bar{\nu}$ mode.}

\end{figure}

\begin{figure}[!h]
\begin{centering}
\begin{tabular}{cc}
\includegraphics[width=0.5\columnwidth]{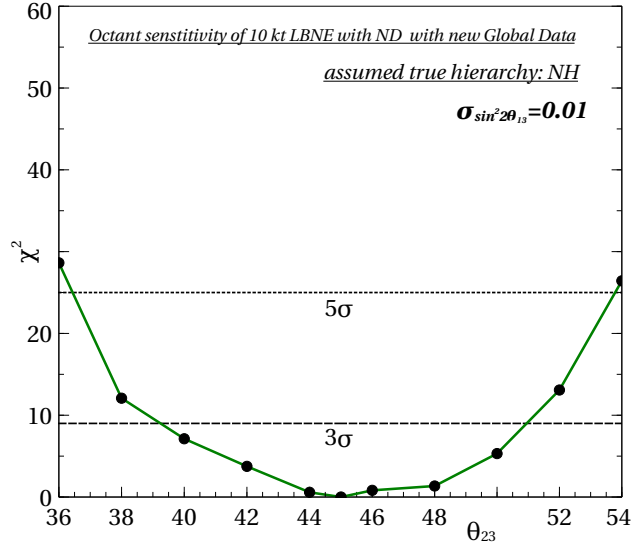}
\end{tabular}
\par\end{centering}
\caption{Octant sensitivity plot for LBNE (10 kt FD with ND) with recent global data. Comparing this plot with the green solid plot of fig. 1 (for NH), it is observed that octant sensitivity with new global data decreases slightly at $3\sigma$ cl when assumed true hierarchy is NH .  }

\end{figure}

\begin{figure}[!t]
\begin{centering}
\begin{tabular}{cc}
\includegraphics[width=0.5\columnwidth]{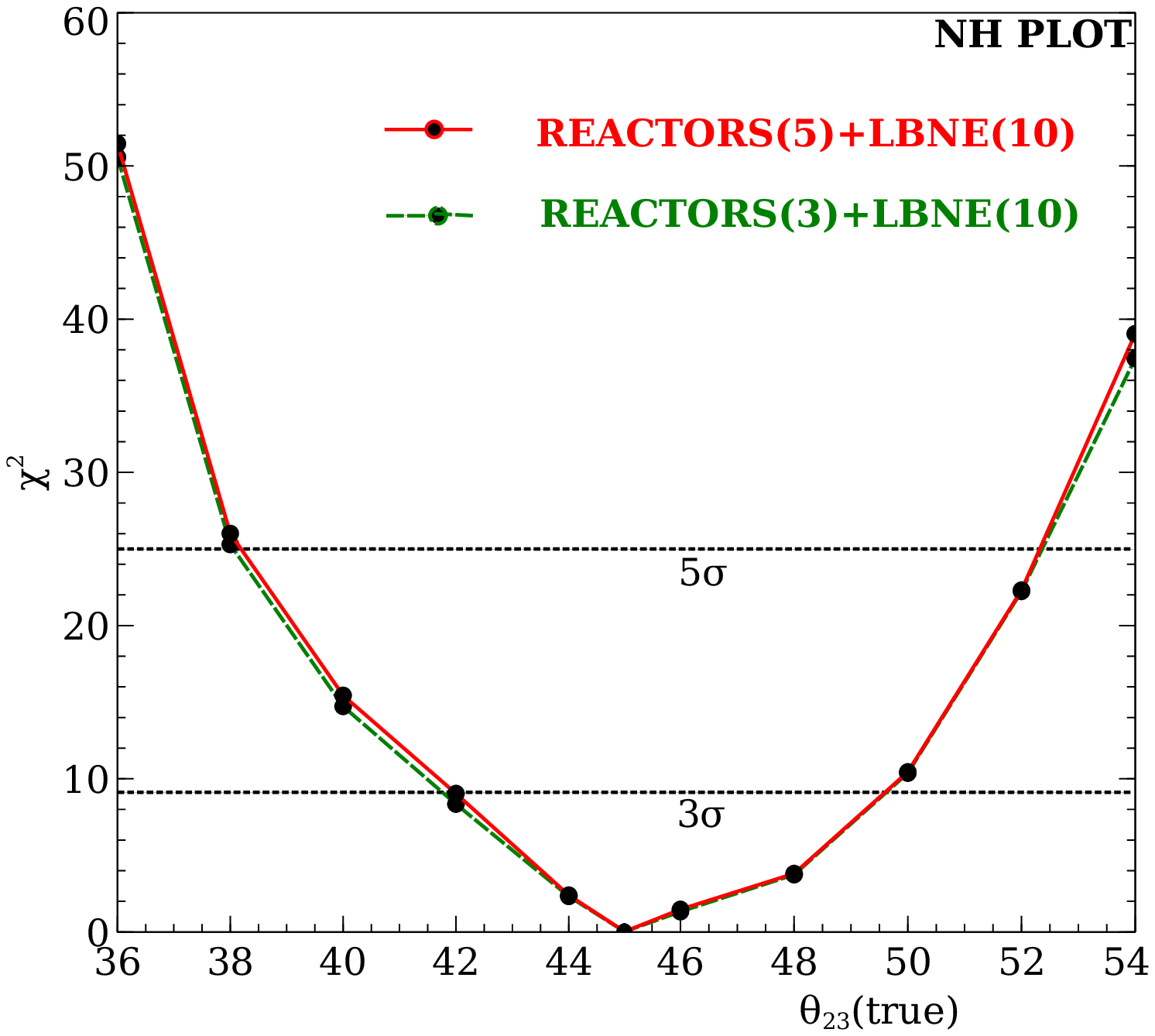}
\includegraphics[width=0.5\columnwidth]{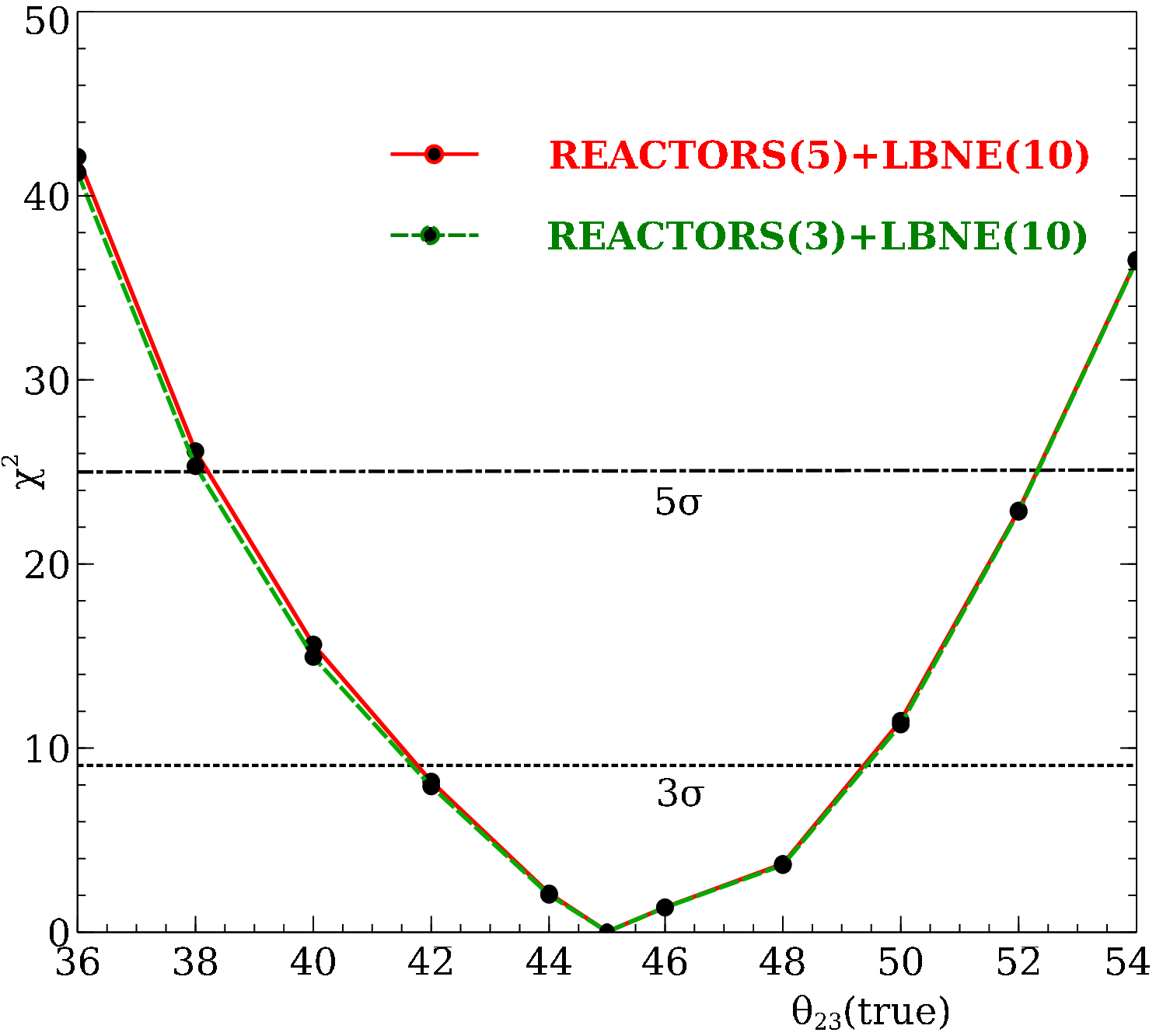}
\end{tabular}
\par\end{centering}

\caption{Octant sensitivity plot for the combined set-up (i.e. DB+DC+RENO+LBNE) with different exposure both in NH (1st plot) and IH (2nd plot) mode with ND. The Red plot (Green dashed plot) is obtained by combining 5 years (3 years) data of each reactor with 10 (5 years in $\nu$ and 5 years in $\bar{\nu}$ mode) years LBNE data. The difference between the plots is negligible i.e. increasing the exposure of reactor experiments does not affect the octant sensitivity of LBNE noticeably.  }

\end{figure}

\begin{figure}[!h]
\begin{centering}
\begin{tabular}{cc}
\includegraphics[width=0.5\columnwidth]{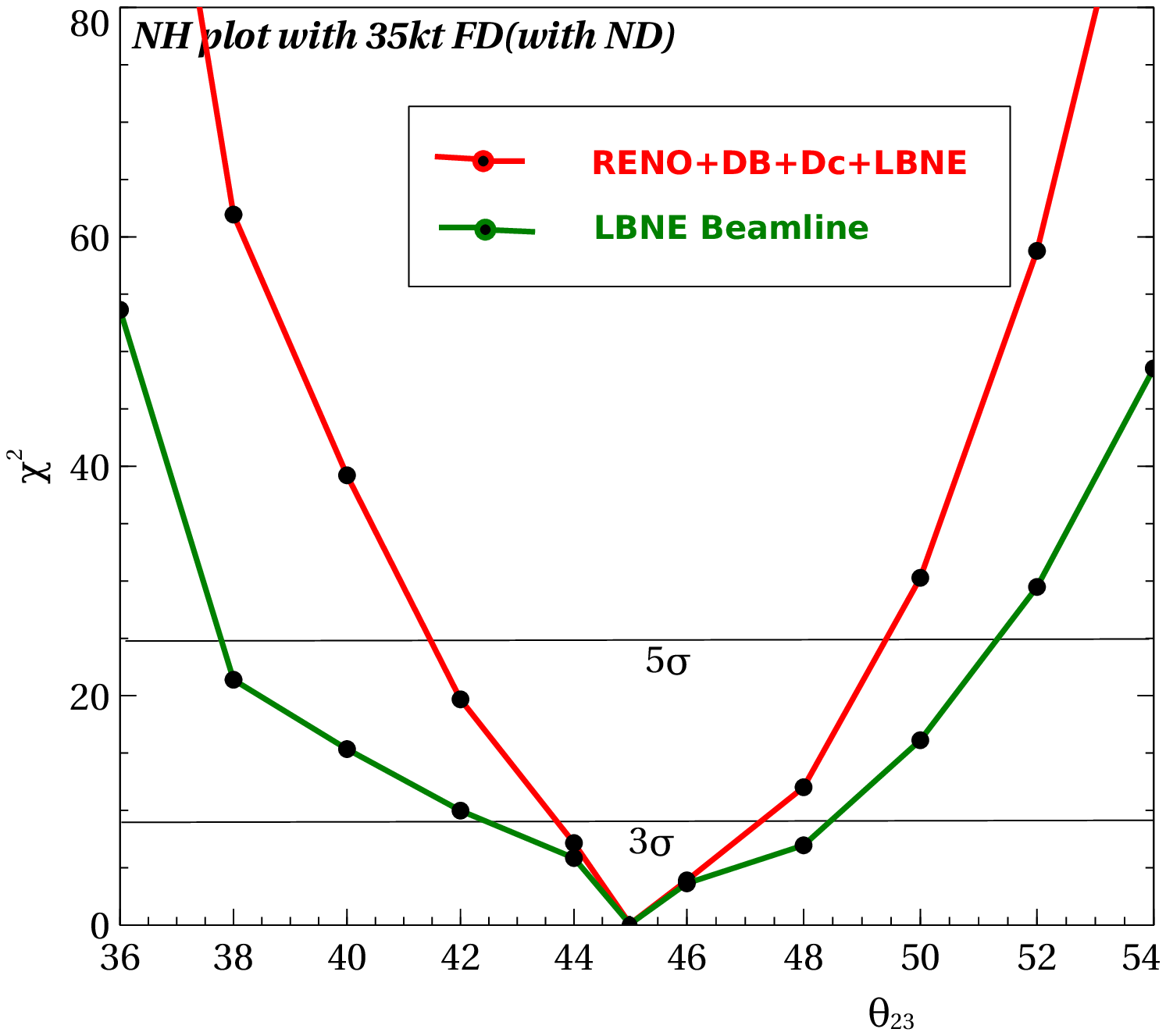}
\includegraphics[width=0.5\columnwidth]{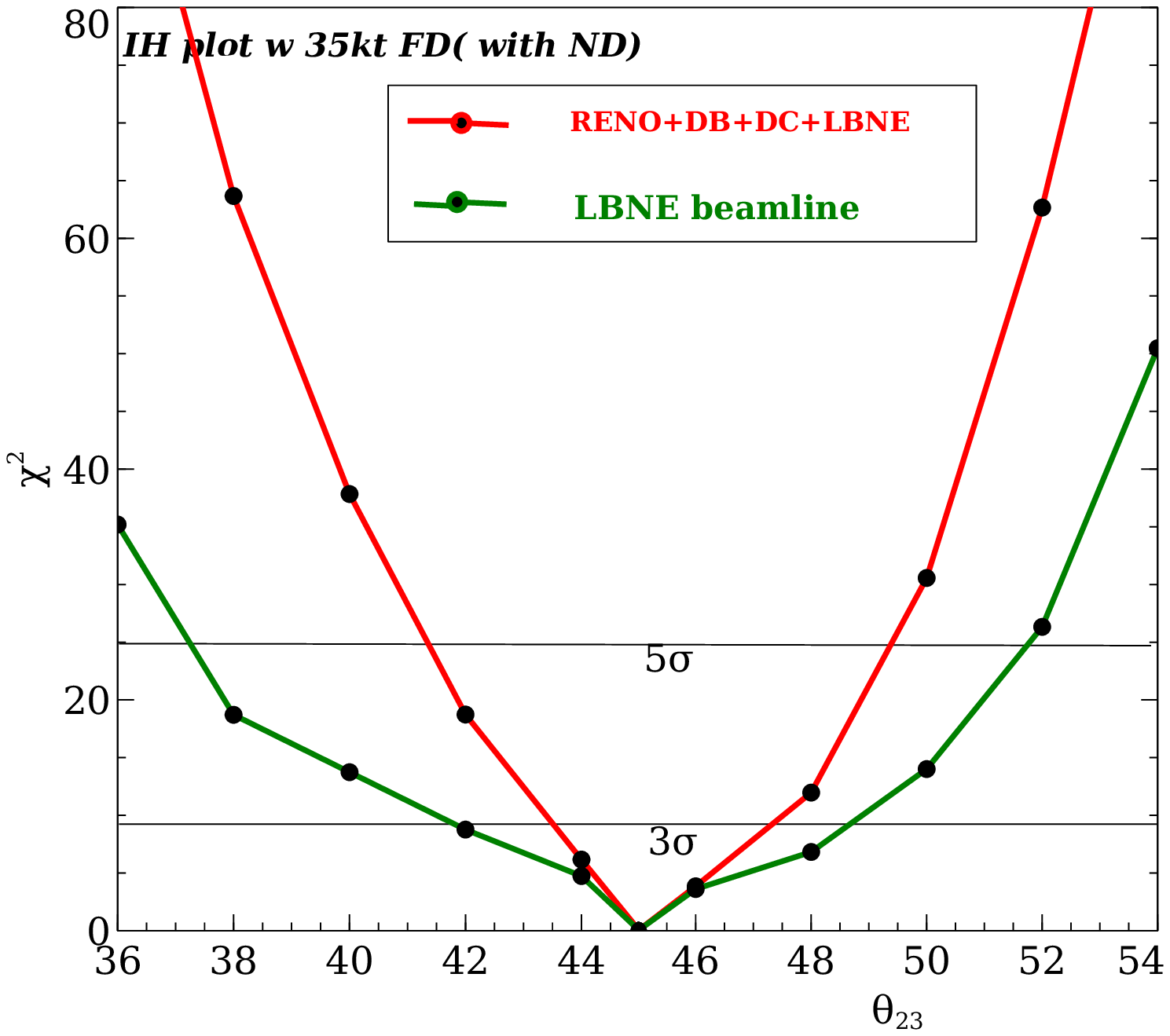}\\
\includegraphics[width=0.5\columnwidth]{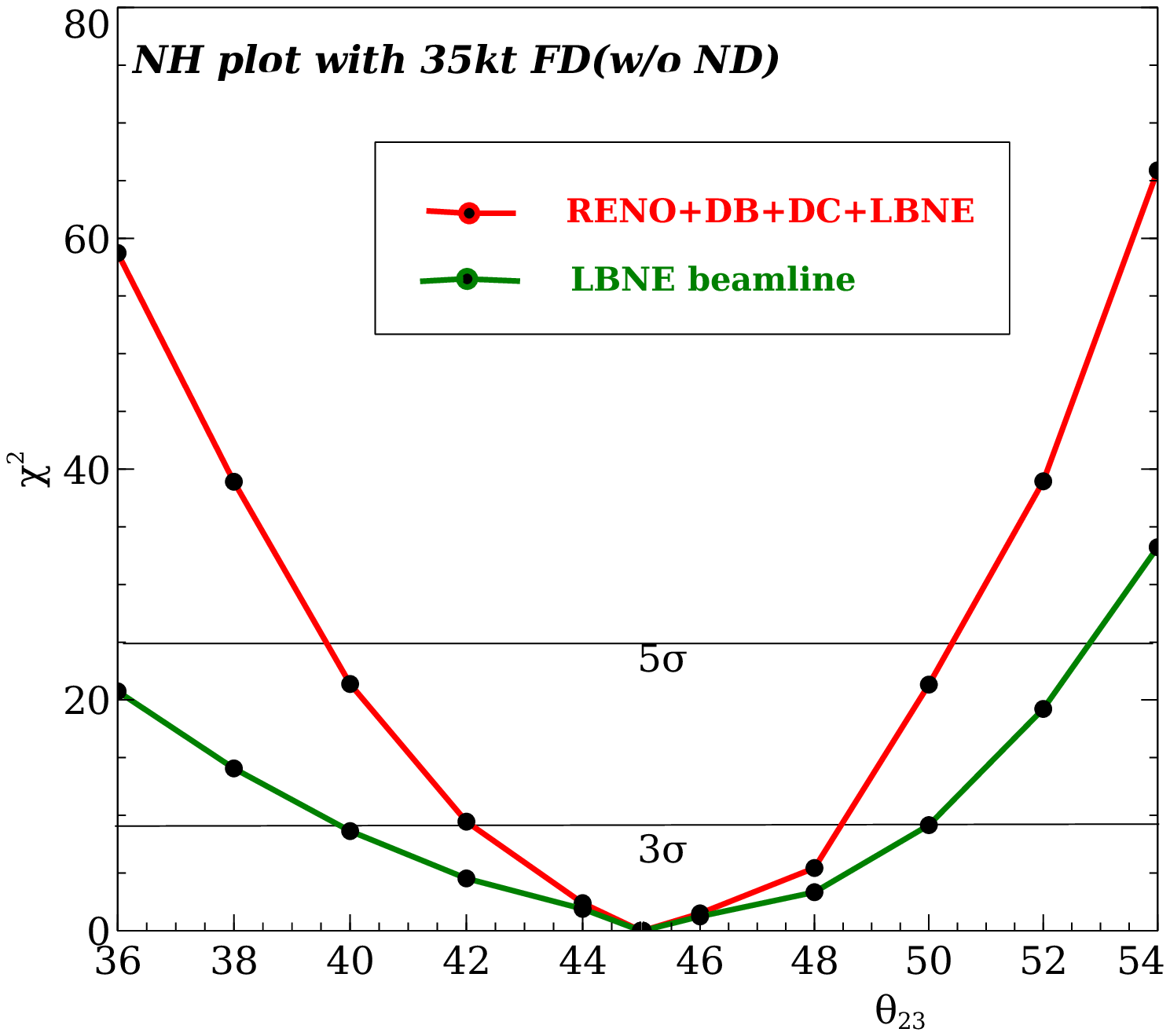}
\includegraphics[width=0.5\columnwidth]{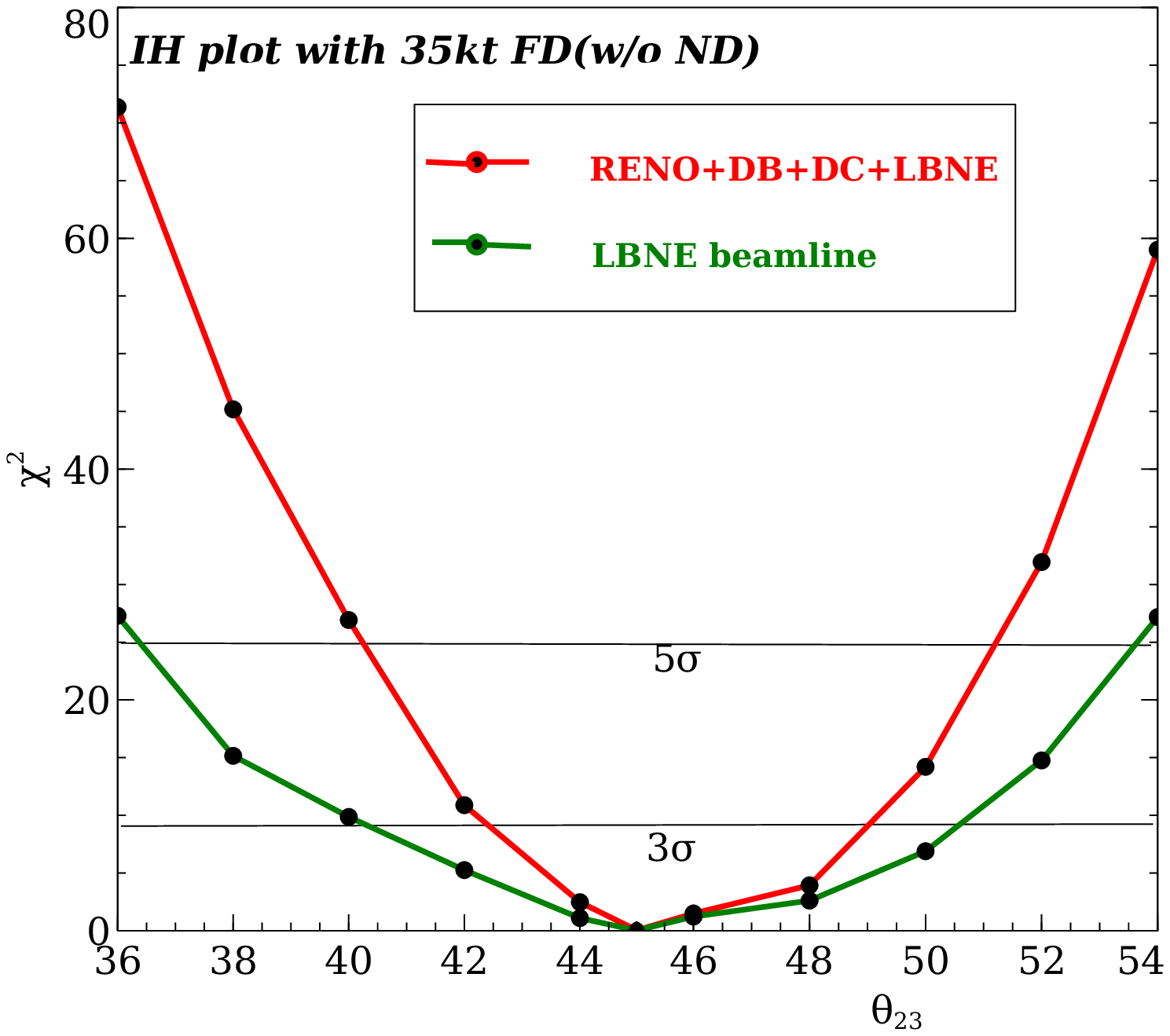}
\end{tabular}
\par\end{centering}
\caption{Octant sensitivity plot for the 35 kt FD with ND (upper panel) and without ND (lower panel) and with prior $\sigma_{\sin^22\theta_{13}} = 0.01$. We have combined 3 years of $\bar{\nu}$ data from each reactor experiments with the (5+5) years of LBNE data in $\nu$ and $\bar{\nu}$ mode with and without any ND for LBNE. }

\end{figure}

 Our test parameters are $\delta_{cp}$, $\bigtriangleup m^2_{31}$ and $\theta_{13}$. We have marginalized over the test parameters $\delta_{cp} \in [-\pi,$ $ \pi]$, $\theta_{13} \in [3^0,$ $11^0]$ and $|\bigtriangleup m^2_{31}| \in [2.19,$ $2.62]\times 10^{-3} eV^2$ range. While examining the octant sensitivity in the lower octant (LO), for any true value of $\theta_{23}$ in the lower octant, we vary the $\theta_{23}$ test parameter in the higher octant in the range $[45^0,$ $ 54^0]$. Similarly, for any true values of $\theta_{23}$ in the HO, test parameter varies in LO in the range $[36^0,$ $45^0]$. $\chi^2$ used for the combined analysis is given as:
         \begin{equation}\chi^2 = \chi^2_{LBNE}+\chi^2_{Reactors}\end{equation}
  While studying octant sensitivity, we cannot neglect the effect of prior information on $\theta_{13}$. The prior on $\sin^22\theta_{13}$ with 1$\sigma$ error range as stated in the literature  of DC, DB and RENO is $\sigma_{\sin^22\theta_{13}}$=0.01 and $\chi^2$ for prior is given as:

\begin{equation}\chi^2_{prior} =( \frac{\sin^22\theta^{true}_{13}-\sin^22\theta_{13}}{\sigma(\sin^22\theta_{13})})^2\end{equation}

We have also shown the octant sensitivity plot for LBNE beamline with an added prior of $\sigma_{\sin^22\theta_{13}} = 0.05\times{\sin^22\theta_{13}}$, which is equivalent to the projected prior of $\sigma_{\sin^22\theta_{13}}=0.005$ in 1$\sigma$ range of $\theta_{13}$. It is also to be noted that we are adding the prior only with LBNE beamline. No prior is added to the combined LBNE + reactors analysis as the reactor data itself gives the information about $\theta_{13}$.

In case of LBNE beamline, $\chi^2$ minimum is calculated as

   \begin{equation} \chi^2_{total}=min(\chi^2_{LBNE}+\chi^2_{prior})\end{equation}
 
In the combined analysis, when we add a reactor to LBNE, minimum $\chi^2$ is calculated as:

   \begin{equation} \chi^2_{total}=min(\chi^2_{LBNE}+\chi^2_{DB/DC/RENO})\end{equation}

Finally, when all the three reactors DC, DB, RENO are added to LBNE together, minimum $\chi^2$ is calculated as:
   
  \begin{equation}\chi^2_{total}=min(\chi^2_{LBNE}+\chi^2_{DC}+\chi^2_{DB}+\chi^2_{RENO})\end{equation}\\

\section{Results}
   
Some abbreviations for the figures are:

LO-NH (IH)-true-NH (IH)-test : We have fixed the true values of $\theta_{23}$ in LO and the true hierarchy is NH (IH). The test hierarchy is also NH (IH).

HO-NH (IH)-true-NH (IH)-test : We have fixed the true values of $\theta_{23}$  in HO and the true hierarchy is NH (IH). The test hierarchy is also NH (IH).

 \begin{figure}[!h]
\begin{centering}
\begin{tabular}{cc}
\includegraphics[width=0.52\columnwidth]{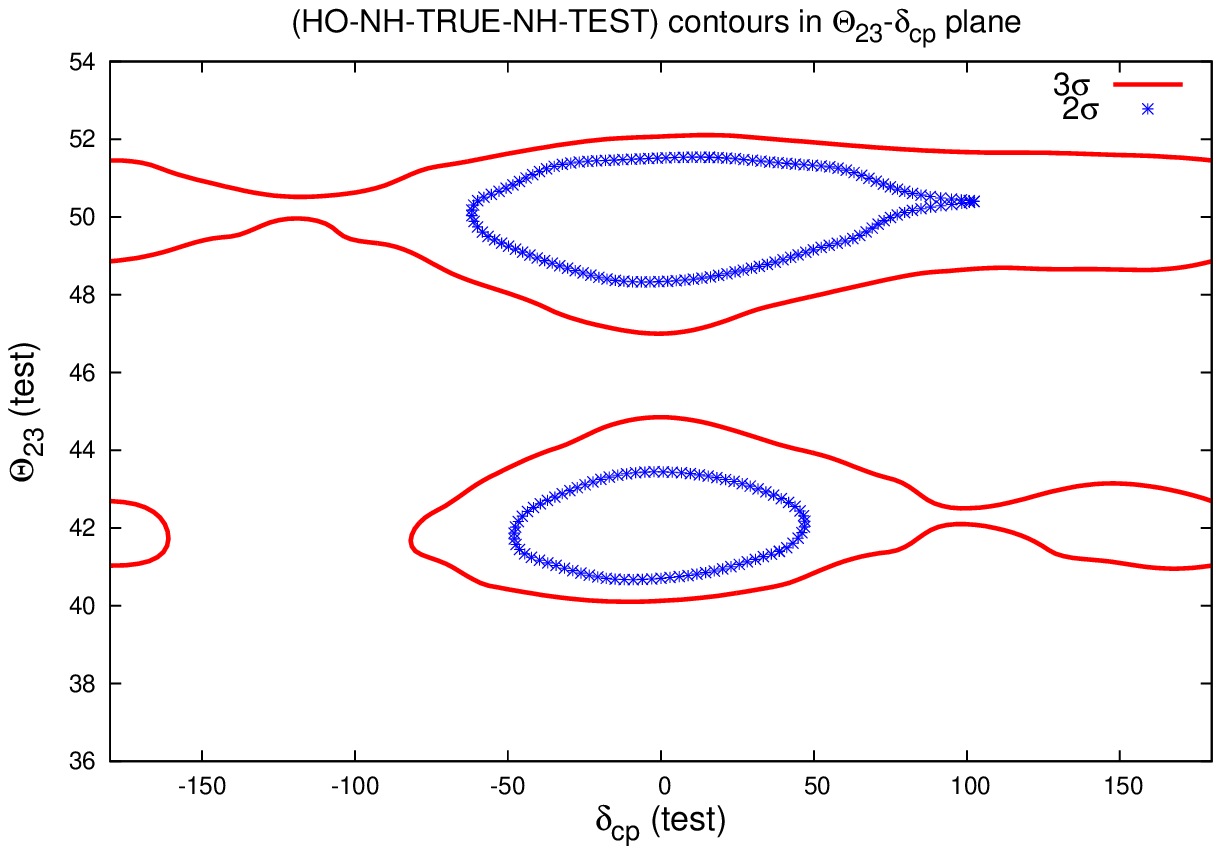}
\includegraphics[width=0.52\columnwidth]{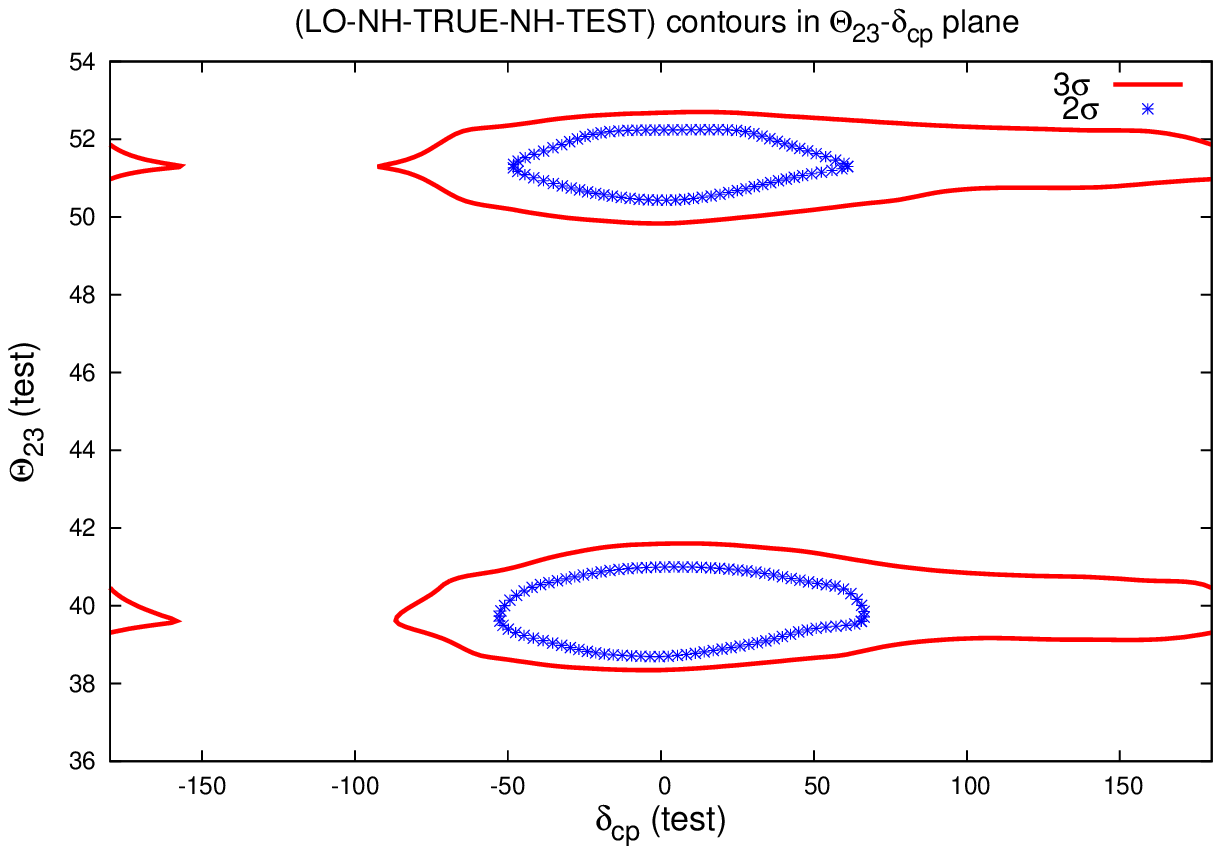}\\
\includegraphics[width=0.52\columnwidth]{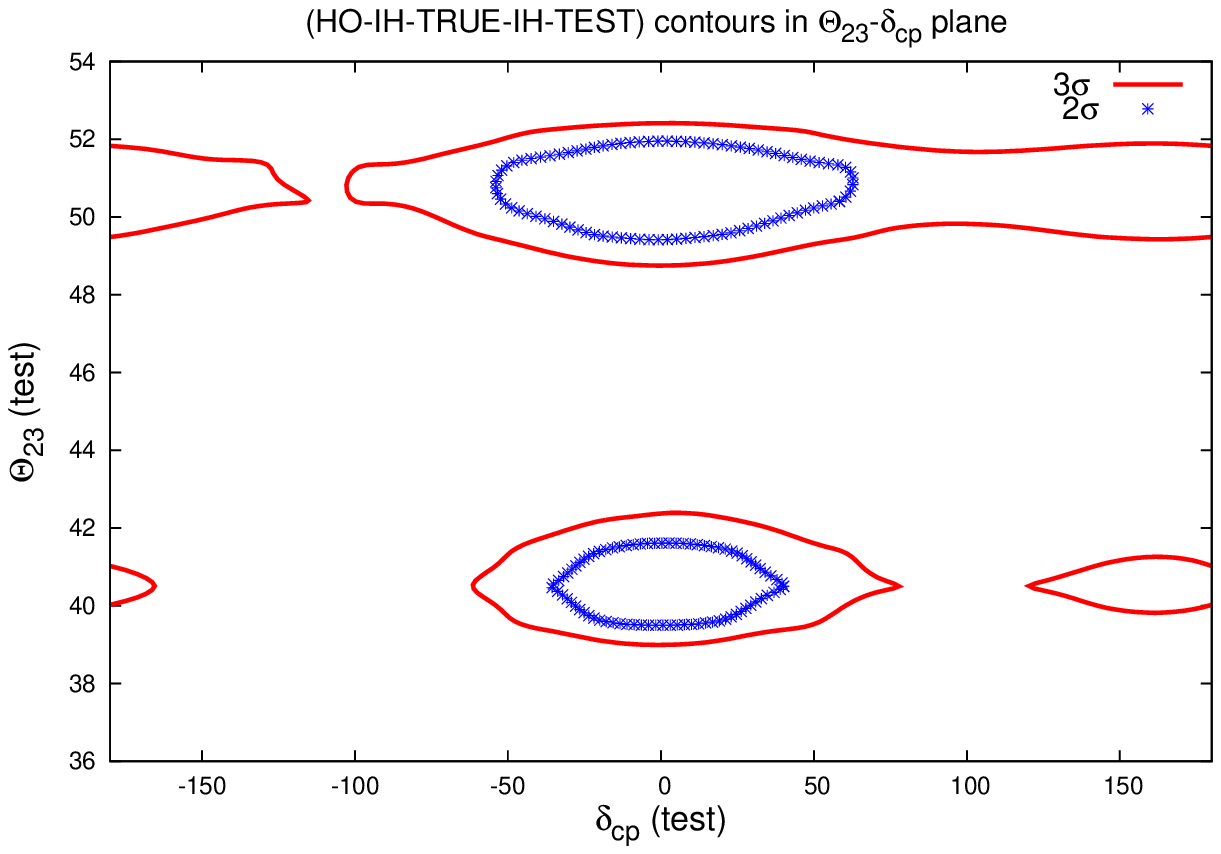}
\includegraphics[width=0.52\columnwidth]{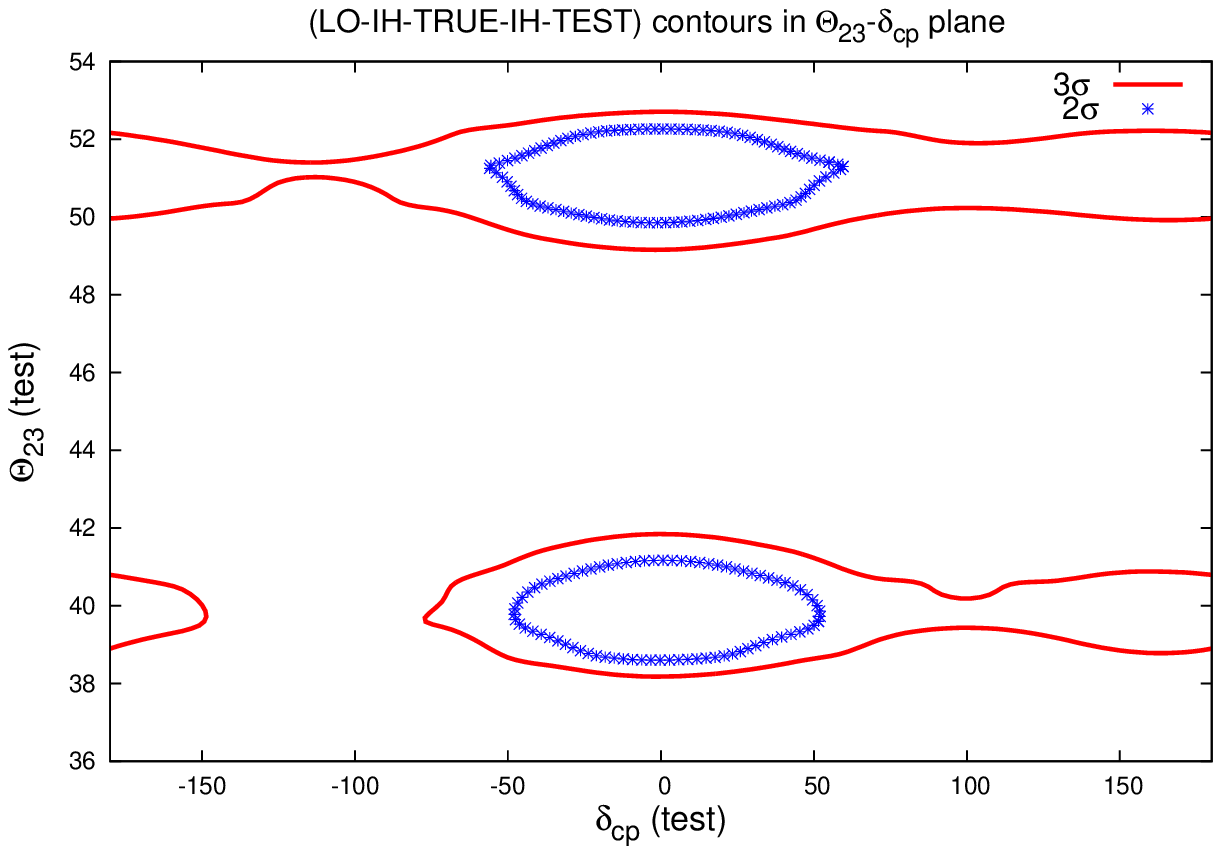}
\end{tabular}
\par\end{centering}

\caption{ These results are shown for LBNE beamline (10 kt FD with ND) in $\delta_{cp}(test)- \theta_{23}(test)$ plane ( 2 d.o.f.) for the true value of $\theta_{23}$ in the HO (LO) and the true value of $\delta_{cp}$= 0.0, marginalizing over $\theta_{13}$ and hierarchy. No prior is added in this case. The figures in the upper (lower) panel are for NH (IH) as true hierarchy. Left(right) figures are for HO (LO) as true octant in both the panels. Plots are shown in 2$\sigma$($\chi^2$= 6.18 at 2 d.o.f.) and 3$\sigma$($\chi^2$=11.83 at 2 d.o.f.) cl. }
\end{figure}

 \begin{figure}
\begin{centering}
\begin{tabular}{cc}
\includegraphics[width=0.52\columnwidth]{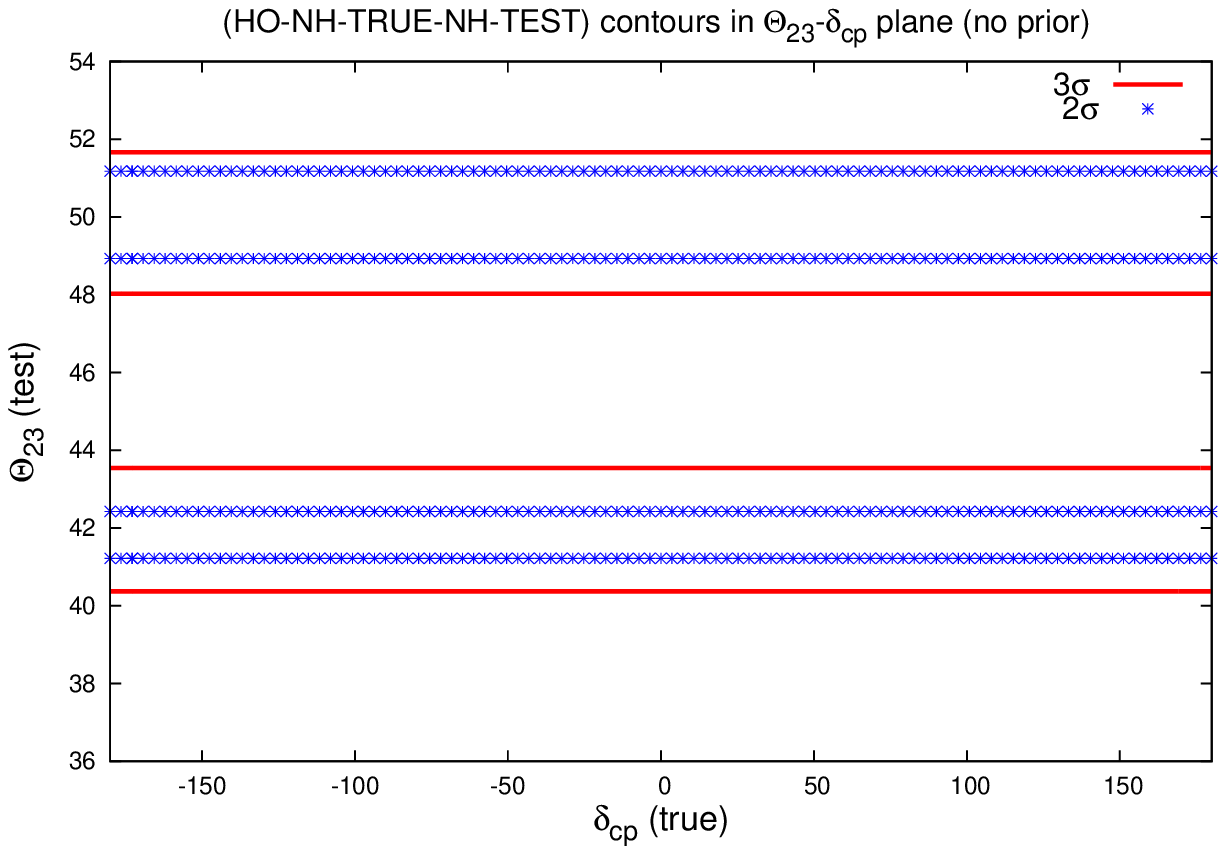}
\includegraphics[width=0.52\columnwidth]{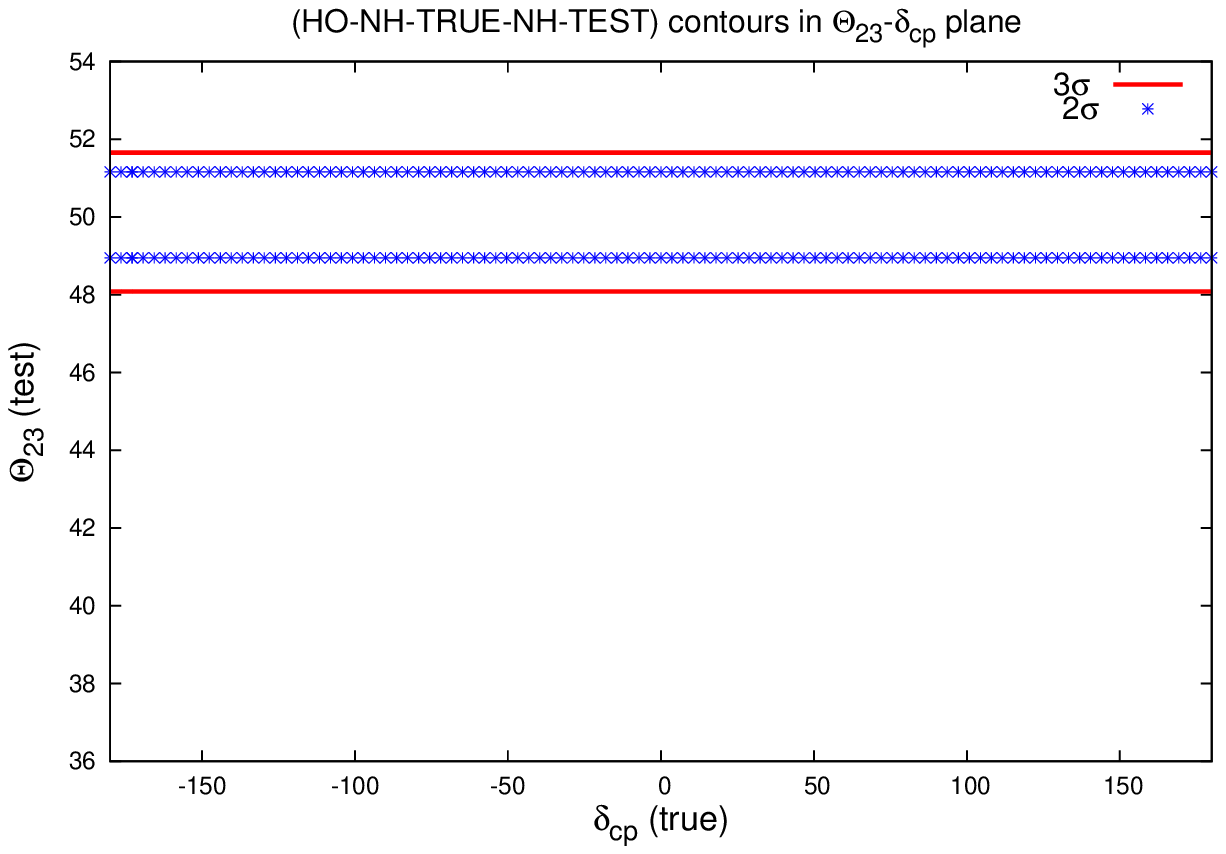}
\end{tabular}
\par\end{centering}

\caption{ Here, we have shown the variation of $\theta_{23}$(test) with $\delta_{cp}$(true) (1 d.o.f.) marginalised over $\theta_{13}$ and $\Delta m^2_{31}$. We have got two bands for LBNE beamline only (10 kt FD with ND) independent of $\delta_{cp}$(true) when we do not add prior
(left figure). But if we include prior on $\theta_{13}$ (i.e. LBNE+ Prior case), (right figure) only one band appears corresponding to HO.  
}

\begin{centering}
\begin{tabular}{cc}
\includegraphics[width=0.52\columnwidth]{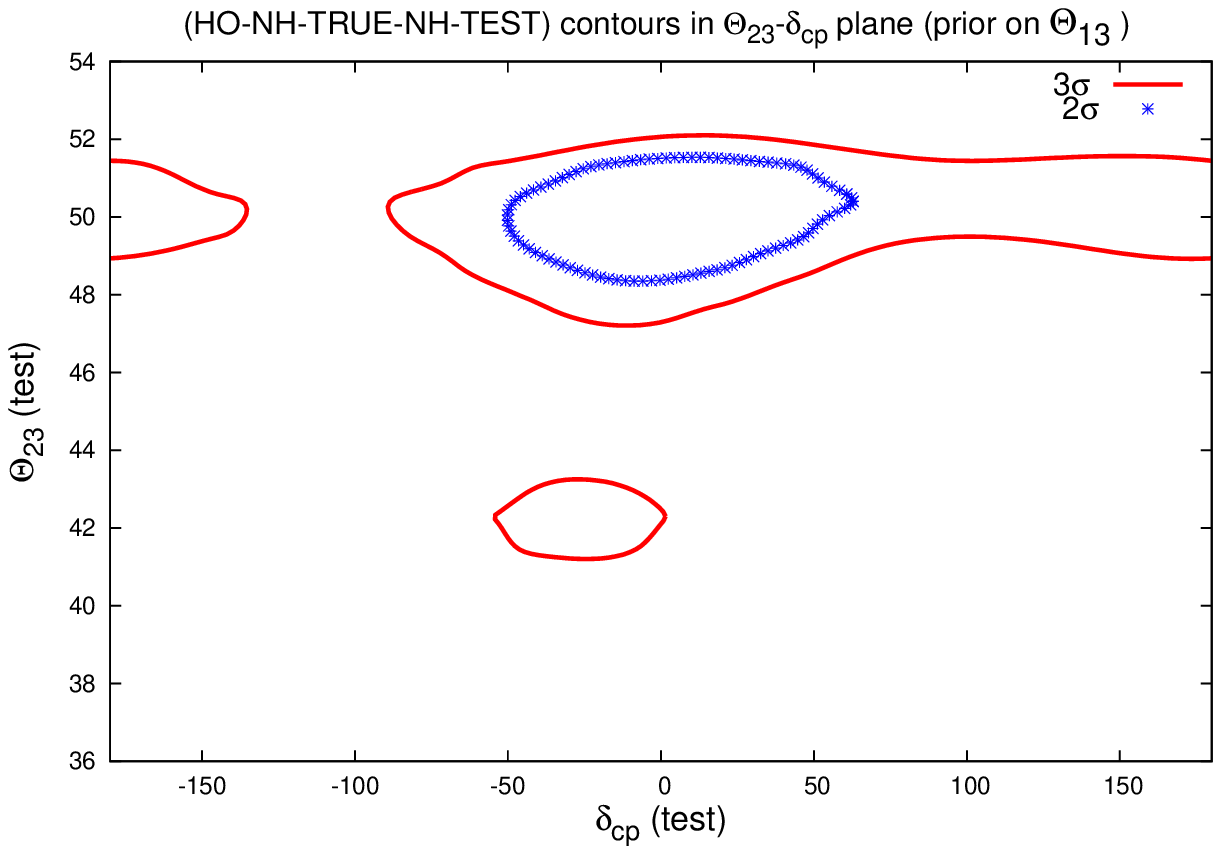}
\includegraphics[width=0.52\columnwidth]{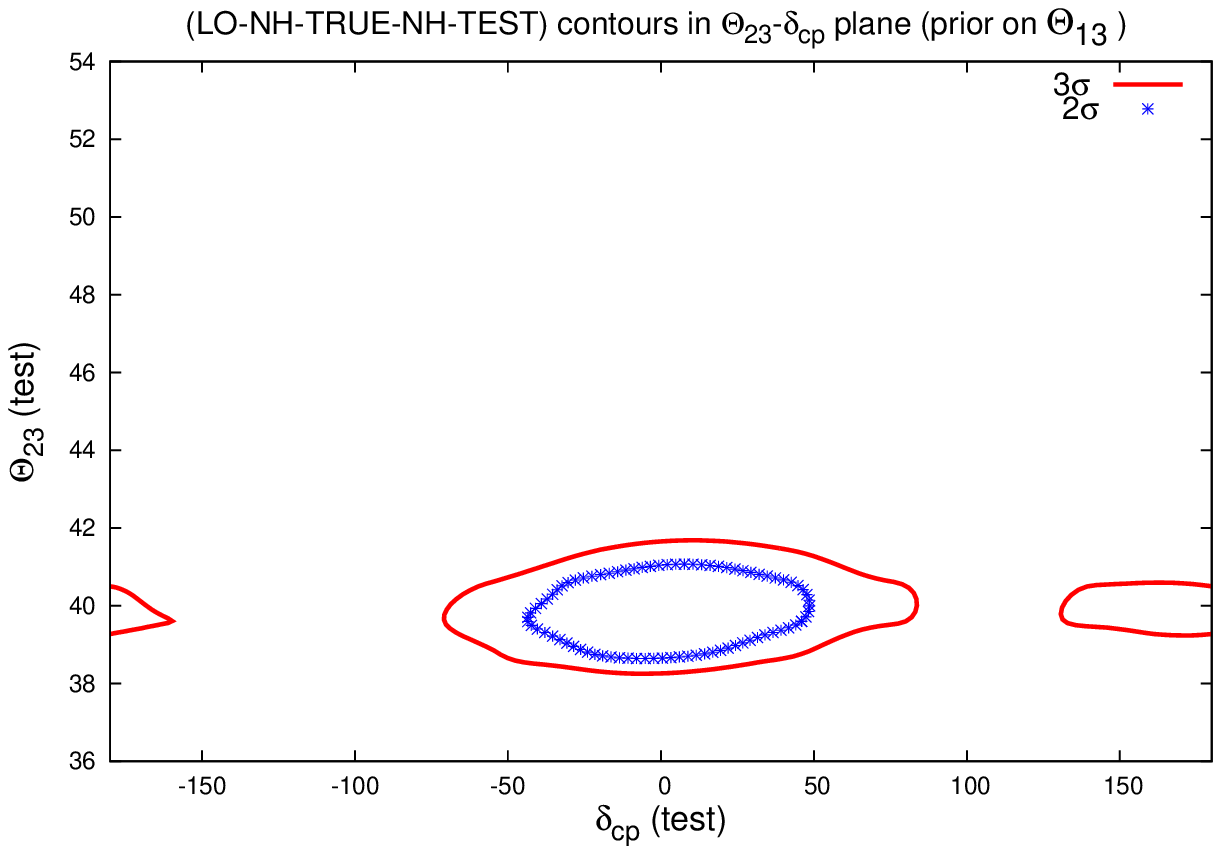}\\
\includegraphics[width=0.52\columnwidth]{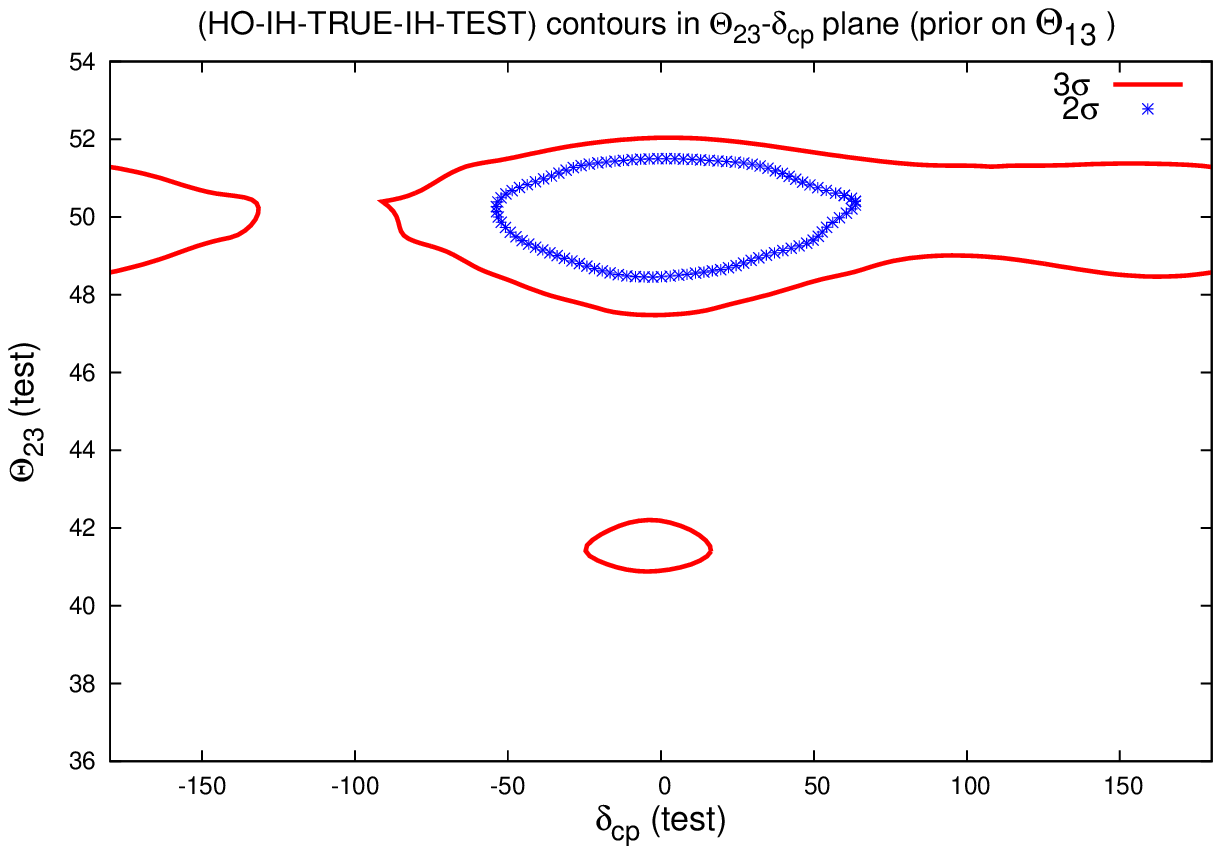}
\includegraphics[width=0.52\columnwidth]{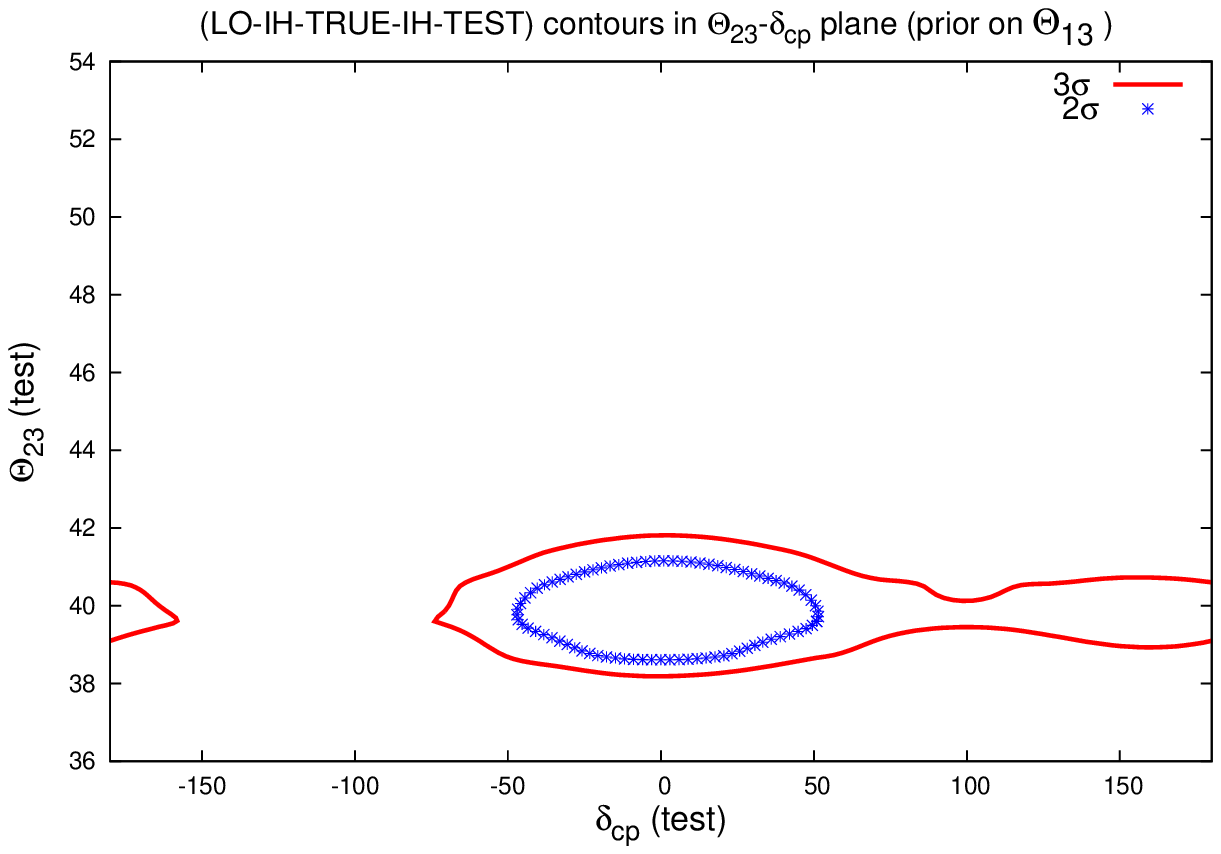}
\end{tabular}
\par\end{centering}

\caption{ Contours are shown for LBNE (10 kt FD with ND)+ prior ($\sigma_{\sin^22\theta_{13}}=0.01$) case, in $\delta_{cp}(test)- \theta_{23}(test)$ plane (2 d.o.f.) for the true value of $\theta_{23}$ in the HO (LO) and the true value of $\delta_{cp}$= 0.0, marginalizing over $\theta_{13}$ and hierarchy. In these plots, we have added prior on $\theta_{13}$. When true value of $\theta_{23}$ is in LO (HO), HO (LO) is ruled out at 2$\sigma$ cl. Plots are shown in 2$\sigma$($\chi^2$= 6.18 at 2 d.o.f.) and 3$\sigma$($\chi^2$= 11.83 at 2 d.o.f.) cl. Figures in the upper (lower) panel are for NH (IH) as true hierarchy. Left (right) figures in both the panel are for HO (LO) as true octant.}

\end{figure}

\begin{figure}[!h]
\begin{centering}
\begin{tabular}{cc}
\includegraphics[width=0.52\columnwidth]{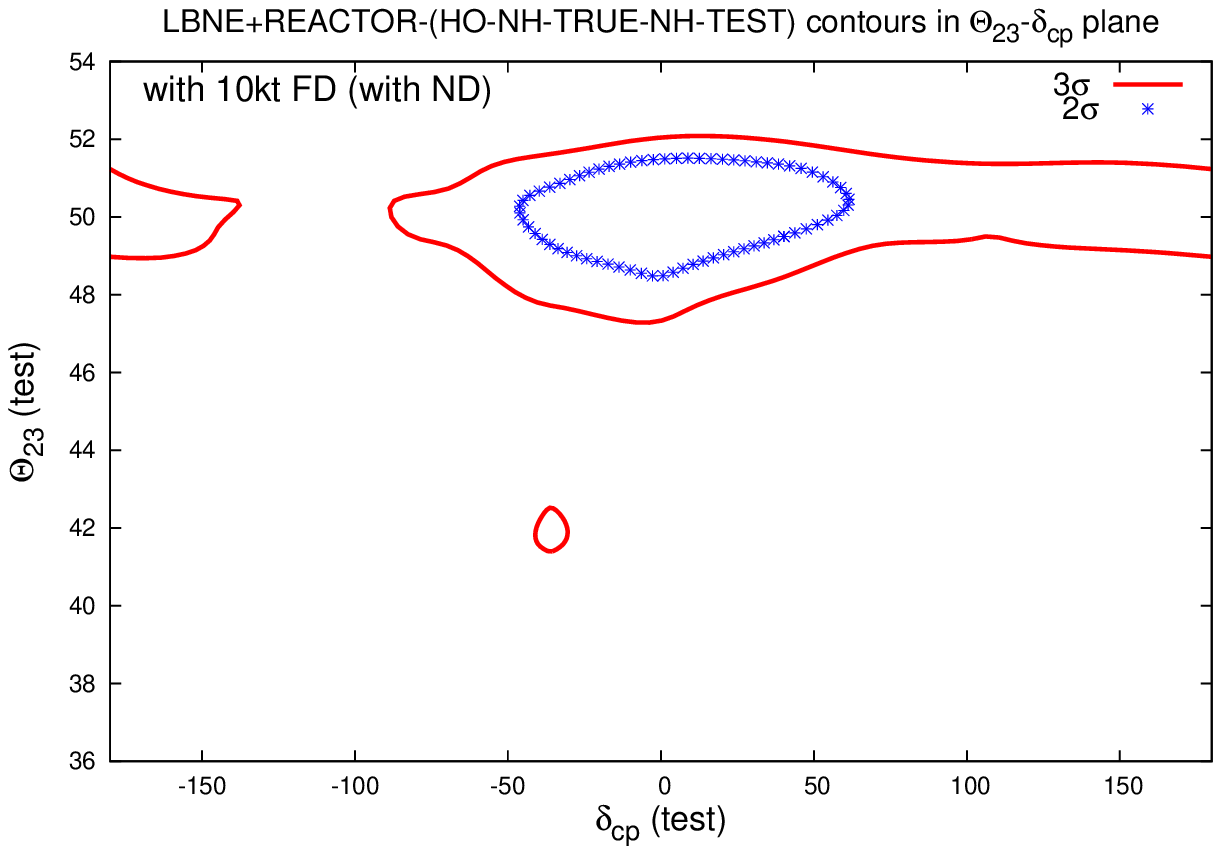}
\includegraphics[width=0.52\columnwidth]{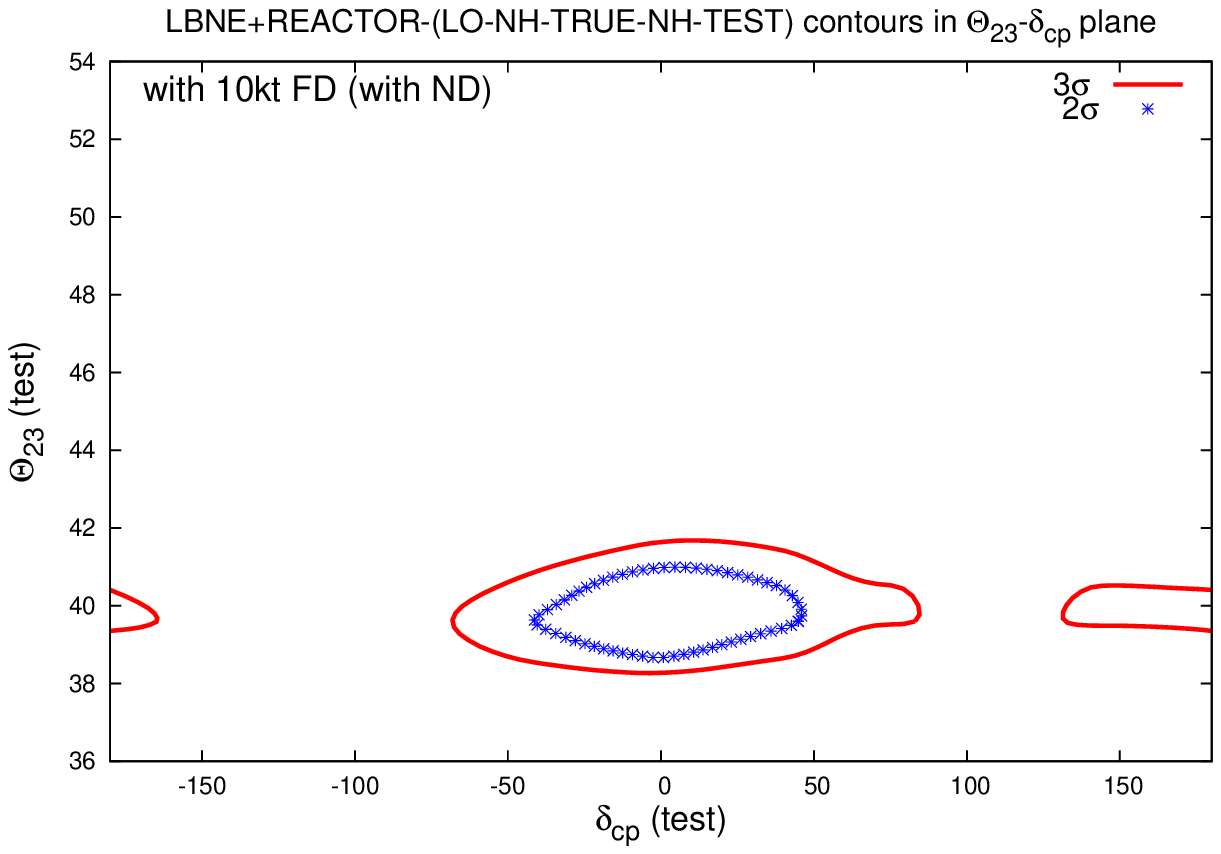}\\
\includegraphics[width=0.52\columnwidth]{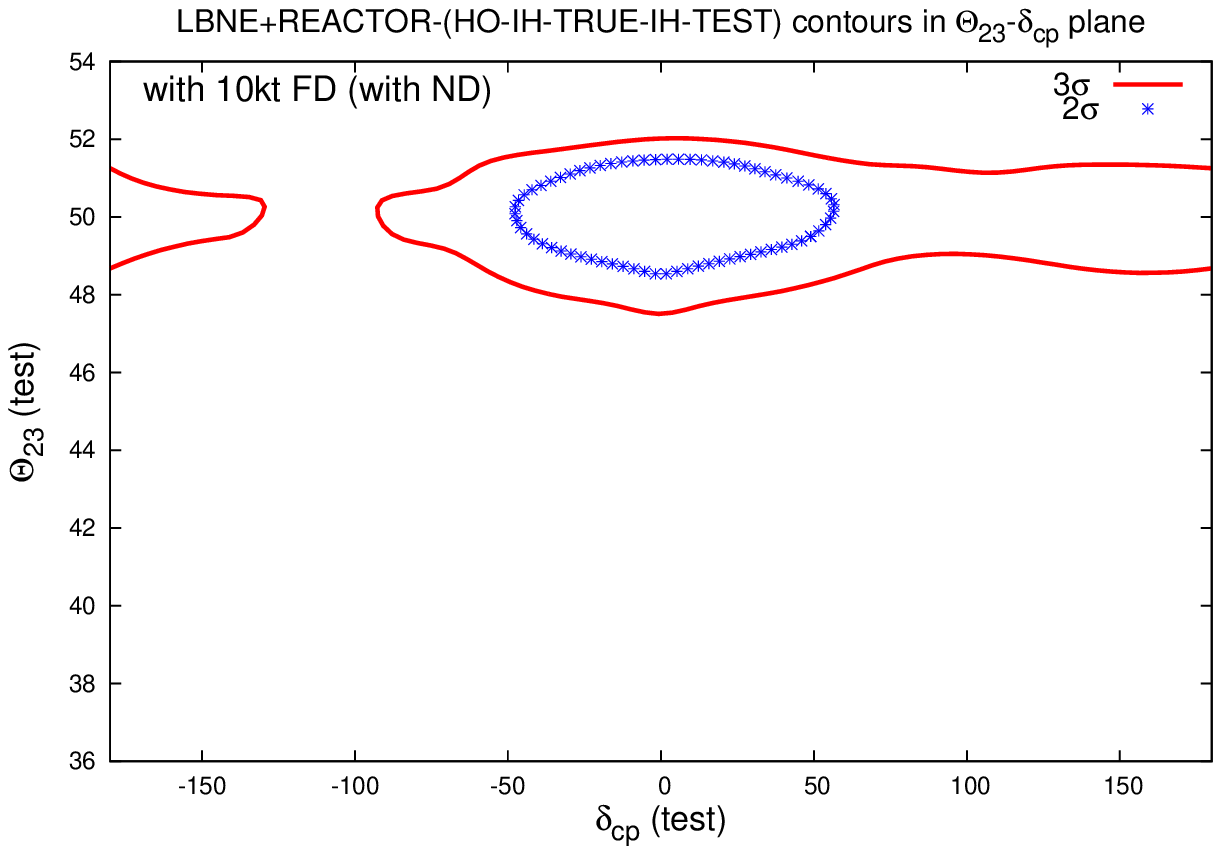}
\includegraphics[width=0.52\columnwidth]{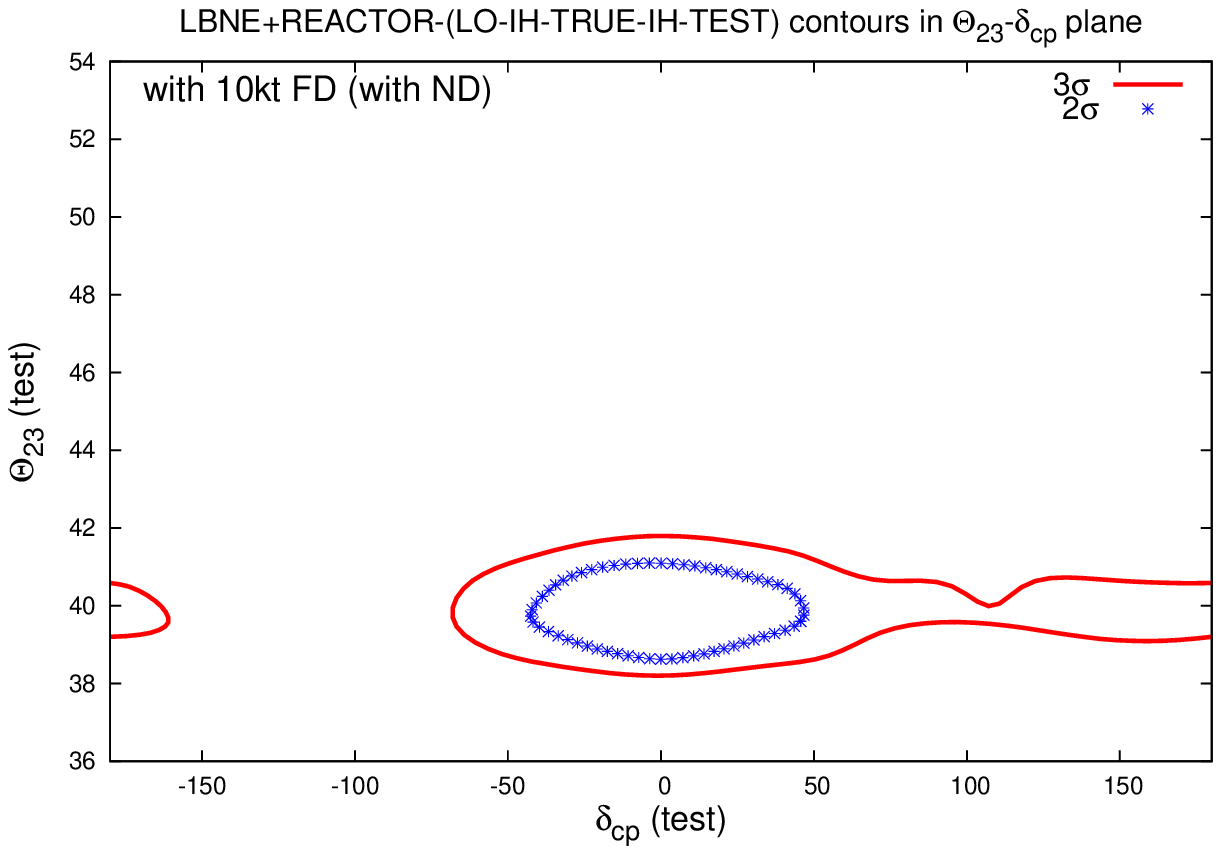}
\end{tabular}
\par\end{centering}

\caption{ Here in both the panels, we have shown our Reactors+LBNE results in 
$\delta_{cp}(test)- \theta_{23}(test)$ plane in 2 d.o.f. for both NH (upper panel) and IH (lower panel). We have combined the 3 years of data from the three reactors with 10 years (5 years for $\nu$ and 5 years for $\bar{\nu}$) of data from LBNE in presence of ND. Prior is not added here. The contours are nearly similar with fig. 5 (LBNE with $\theta_{13}$ prior).  
}

\end{figure}

 \begin{figure}[!h]
\begin{centering}
\begin{tabular}{cc}
\includegraphics[width=0.52\columnwidth]{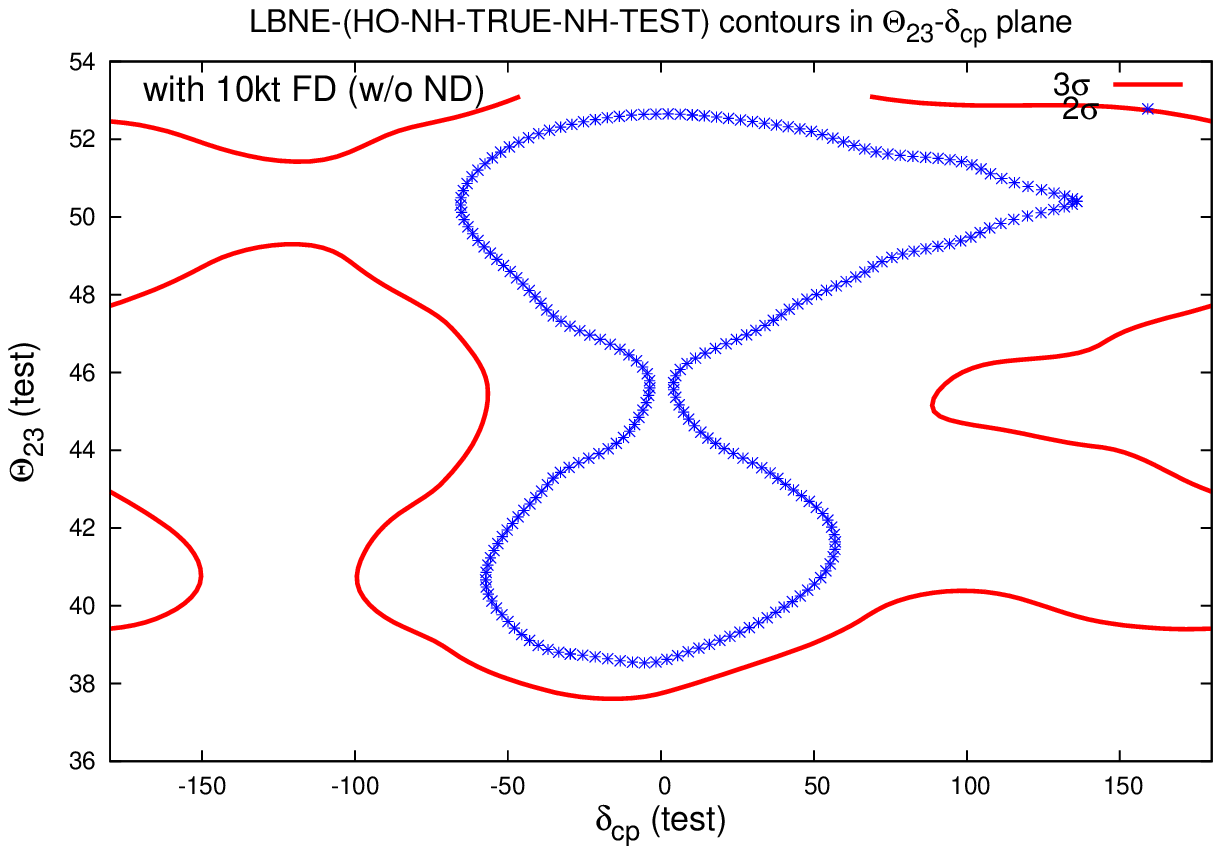}
\includegraphics[width=0.52\columnwidth]{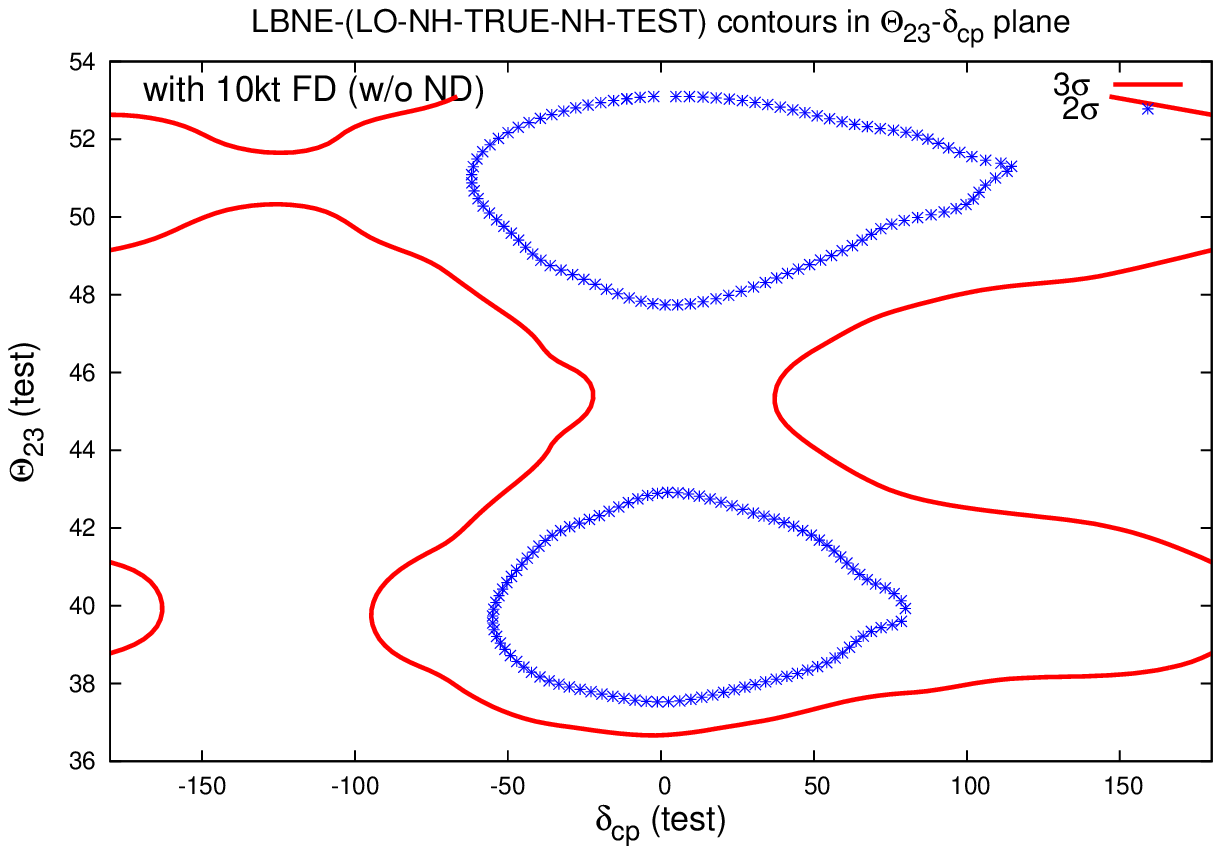}\\
\includegraphics[width=0.52\columnwidth]{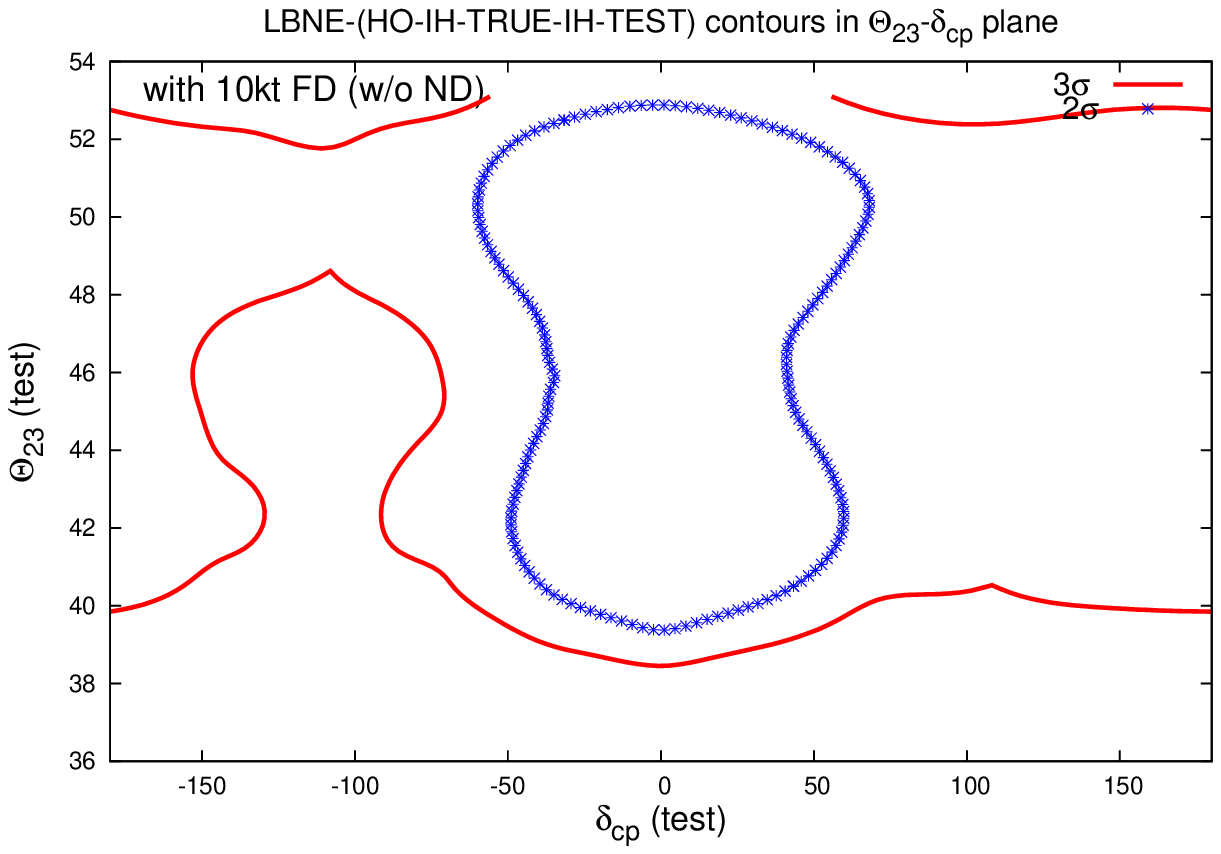}
\includegraphics[width=0.52\columnwidth]{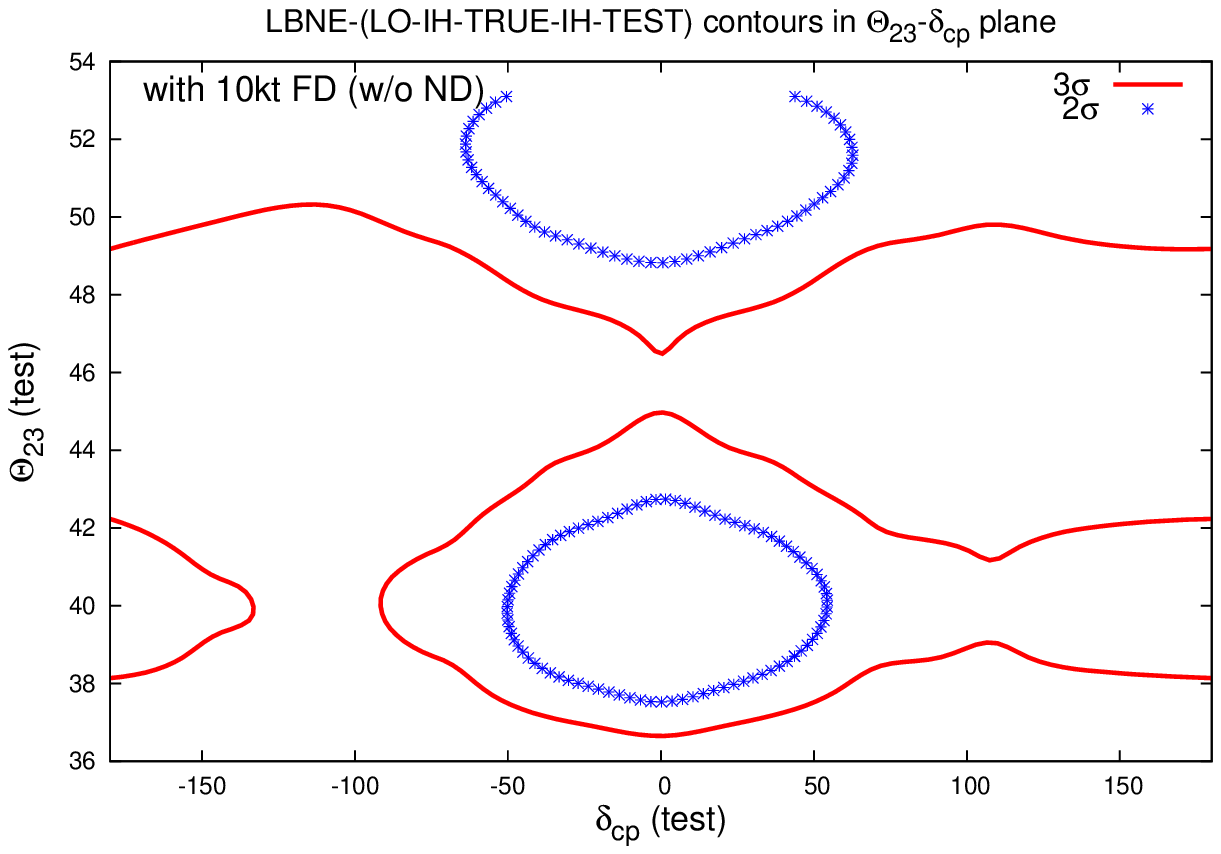}
\end{tabular}
\par\end{centering}

\caption{ These results are shown for LBNE beamline (10 kt FD without ND) in $\delta_{cp}(test)- \theta_{23}(test)$ plane ( 2 d.o.f.) for the true value of $\theta_{23}$ in the HO(LO) and the true value of $\delta_{cp}$= 0.0, marginalizing over $\theta_{13}$ and hierarchy. No prior is added in this case. The figures at the upper (lower) panel are for NH (IH) as true hierarchy. Left (right) figures are for HO (LO) as true octant in both the panel. Plots are shown for 10 years (5 years for $\nu$ and 5 years for $\bar{\nu}$) of LBNE data (without ND) in 2$\sigma$($\chi^2$=6.18 at 2 d.o.f.) and 3$\sigma$($\chi^2$=11.83 at 2 d.o.f.) cl. }
\end{figure}

\begin{figure}[!h]
\begin{centering}
\begin{tabular}{cc}
\includegraphics[width=0.52\columnwidth]{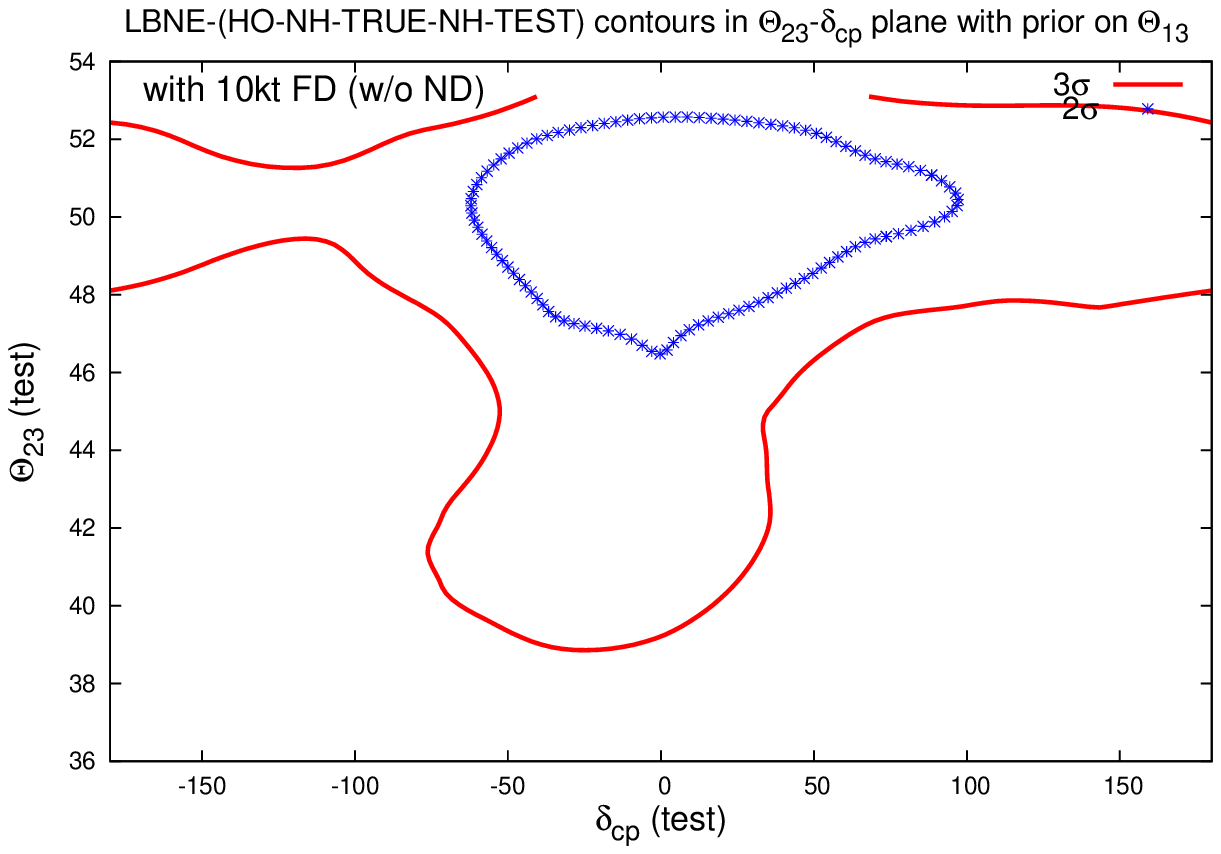}
\includegraphics[width=0.52\columnwidth]{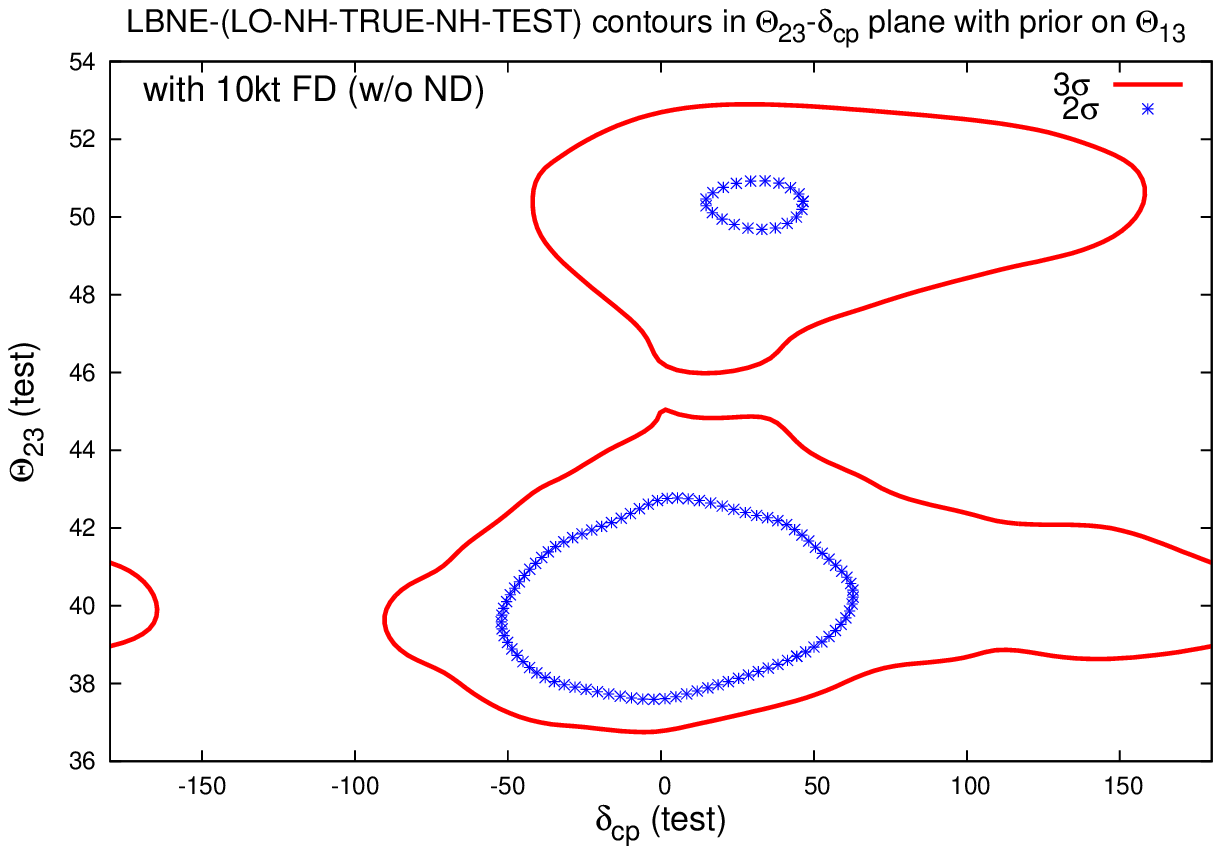}\\
\includegraphics[width=0.52\columnwidth]{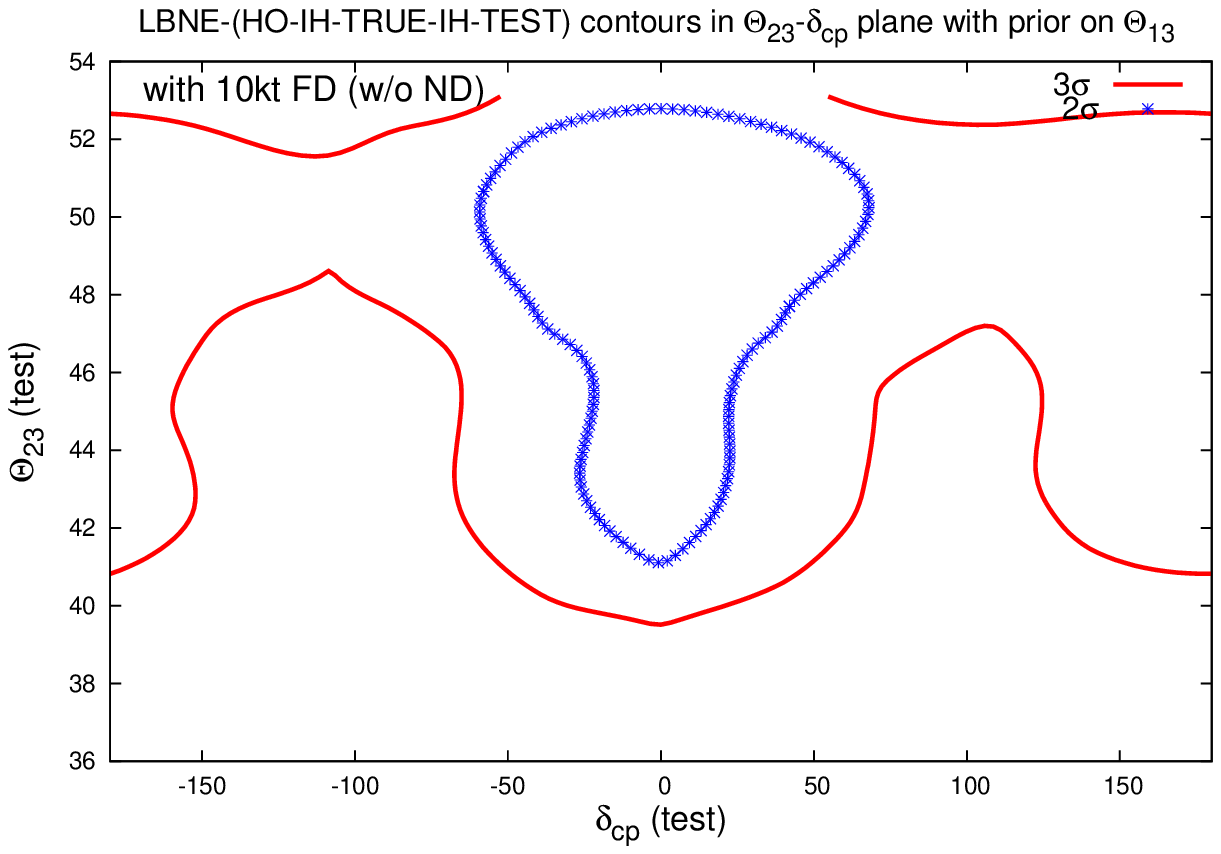}
\includegraphics[width=0.52\columnwidth]{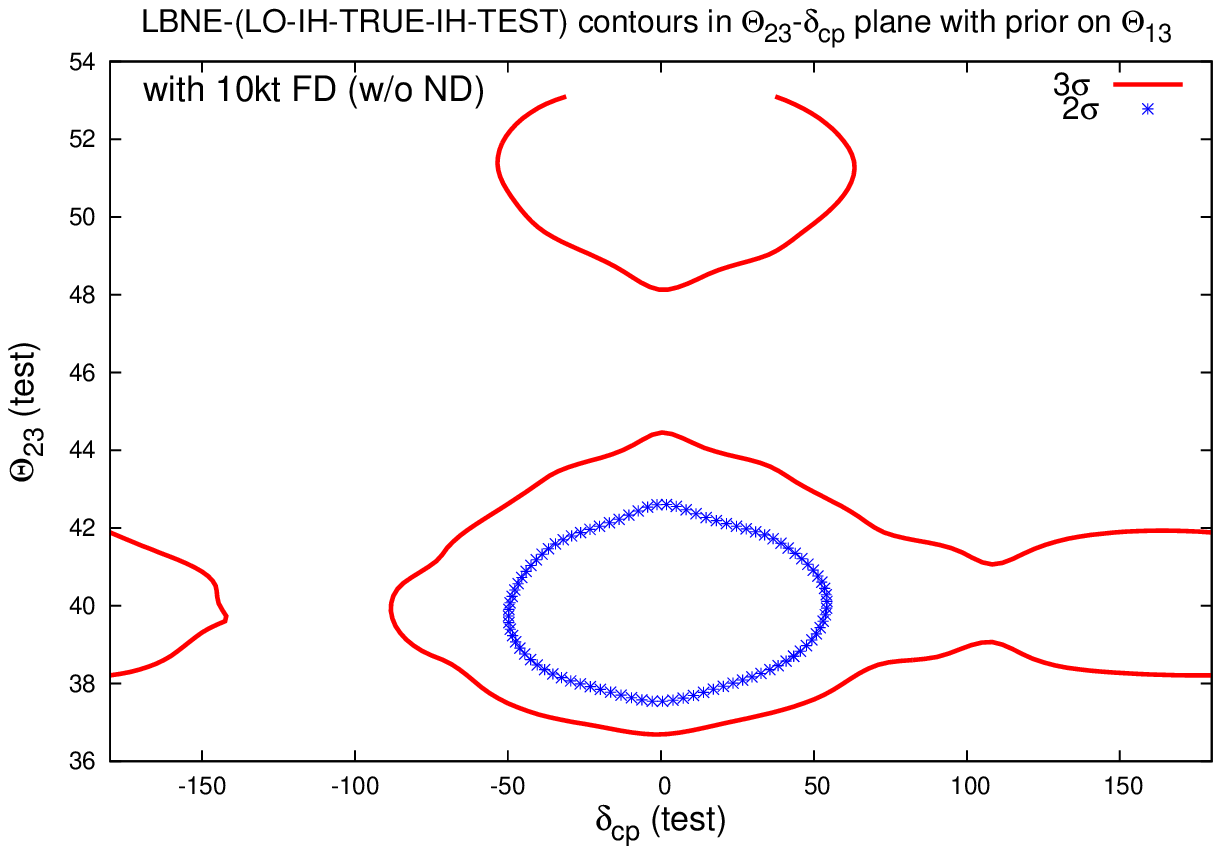}
\end{tabular}
\par\end{centering}

\caption{ Contours are shown for LBNE (10 kt FD without ND)+ prior ($\sigma_{\sin^22\theta_{13}}=0.01$) case, in $\delta_{cp}(test)- \theta_{23}(test)$ plane (2 d.o.f.) for the true value of $\theta_{23}$ in the HO (LO) and the true value of $\delta_{cp}$= 0.0, marginalizing over $\theta_{13}$ and hierarchy. In these plots, we have added prior on $\theta_{13}$. When true value of $\theta_{23}$ is in LO (HO), HO (LO) is ruled out at 2$\sigma$ cl. Figures in the upper (lower) panel are for NH (IH) as true hierarchy. Left (right) figures in both the panel are for HO (LO) as true octant. Plots are shown for 10 years (5 years for $\nu$ and 5 years for $\bar{\nu}$) of LBNE data (without ND) in 2$\sigma$($\chi^2$=6.18 at 2 d.o.f.) and 3$\sigma$($\chi^2$=11.83 at 2 d.o.f.) cl. }

\end{figure}

\begin{figure}[!h]
\begin{centering}
\begin{tabular}{cc}
\includegraphics[width=0.52\columnwidth]{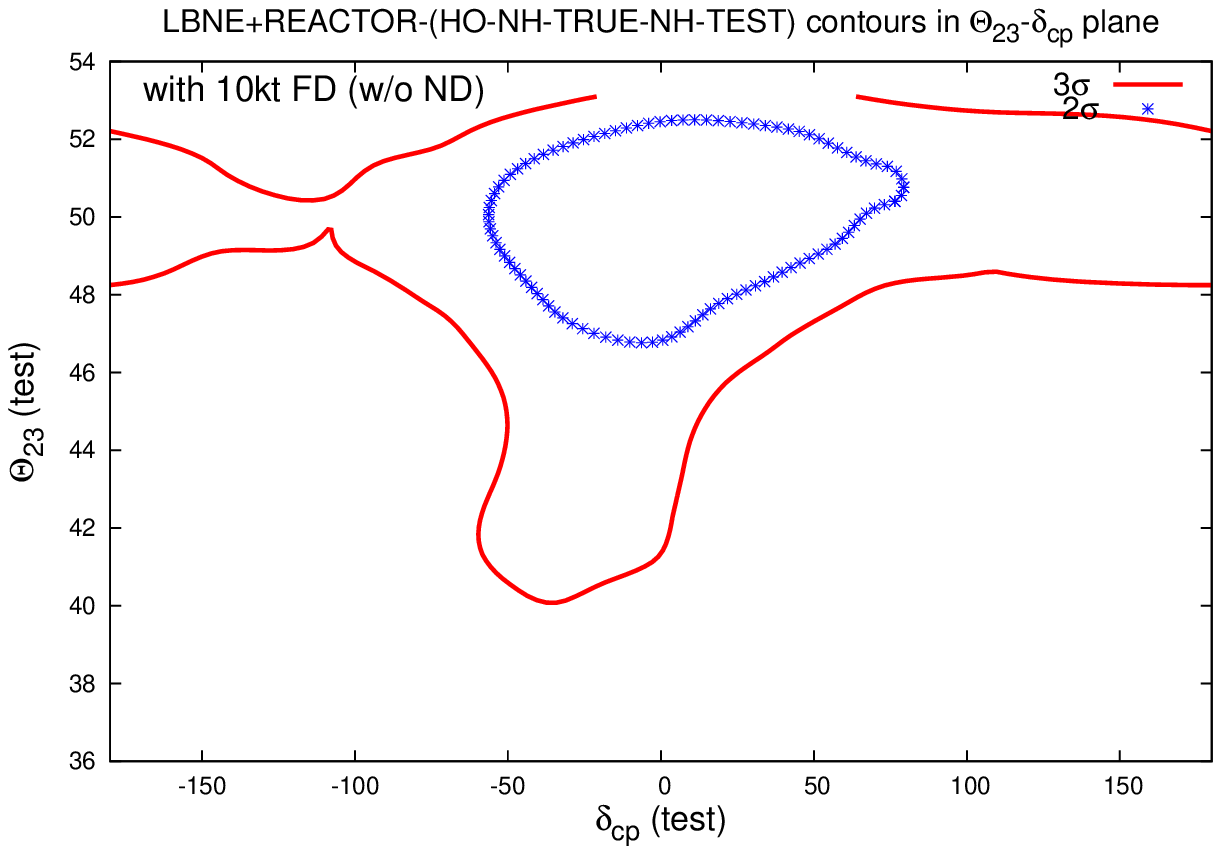}
\includegraphics[width=0.52\columnwidth]{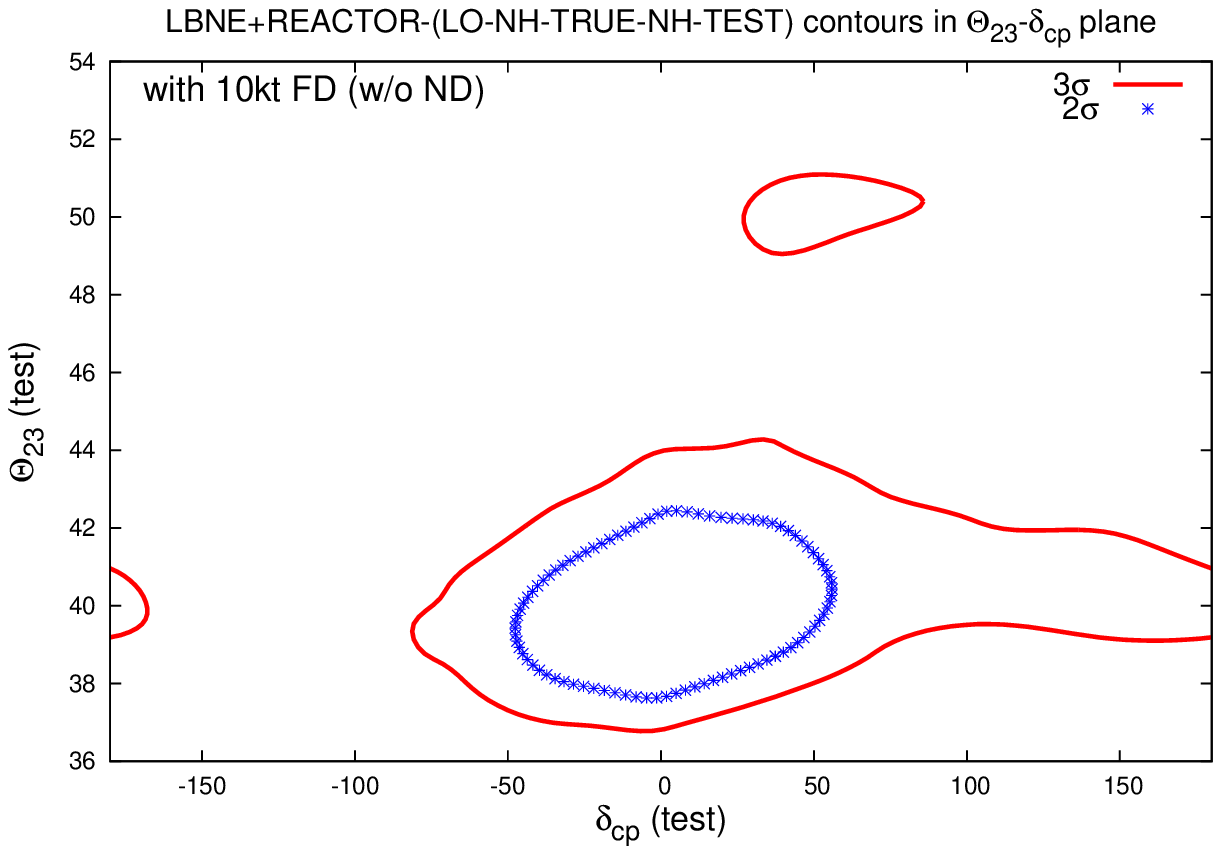}\\
\includegraphics[width=0.52\columnwidth]{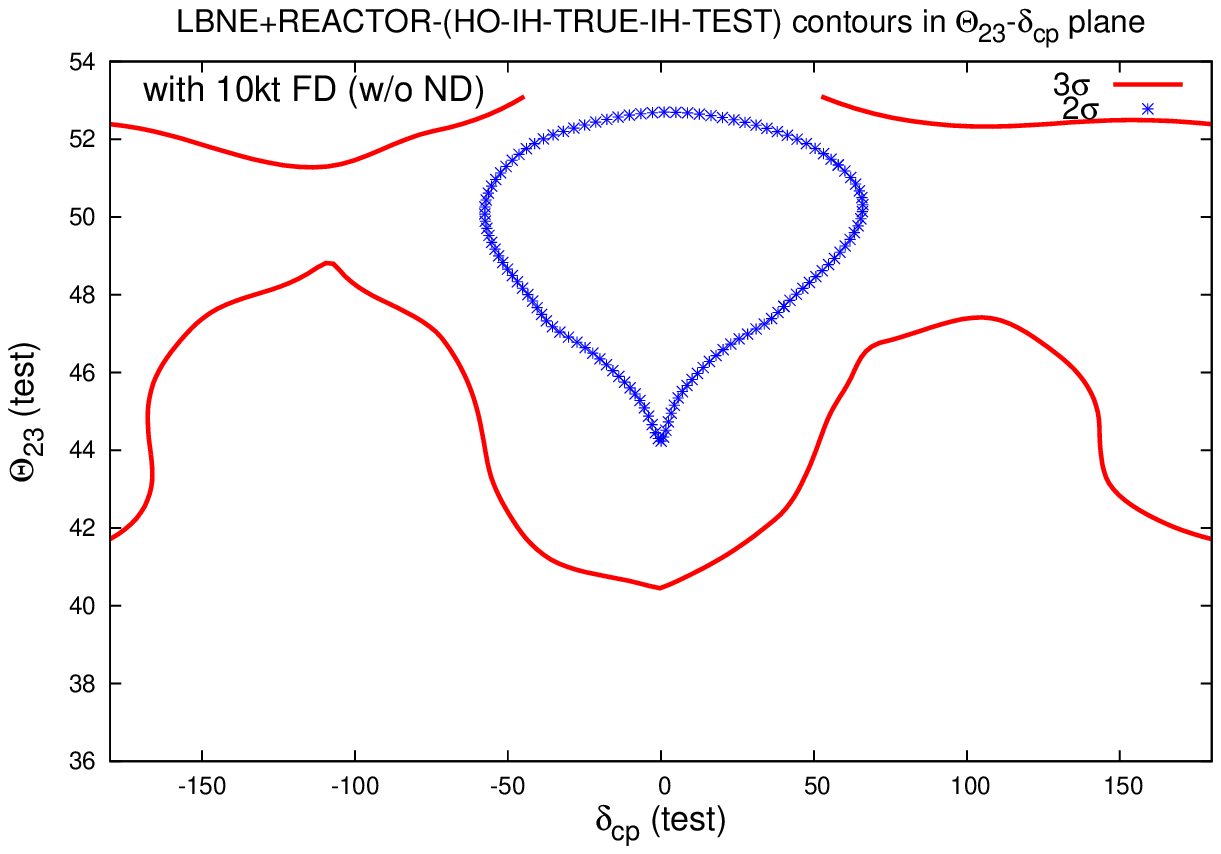}
\includegraphics[width=0.52\columnwidth]{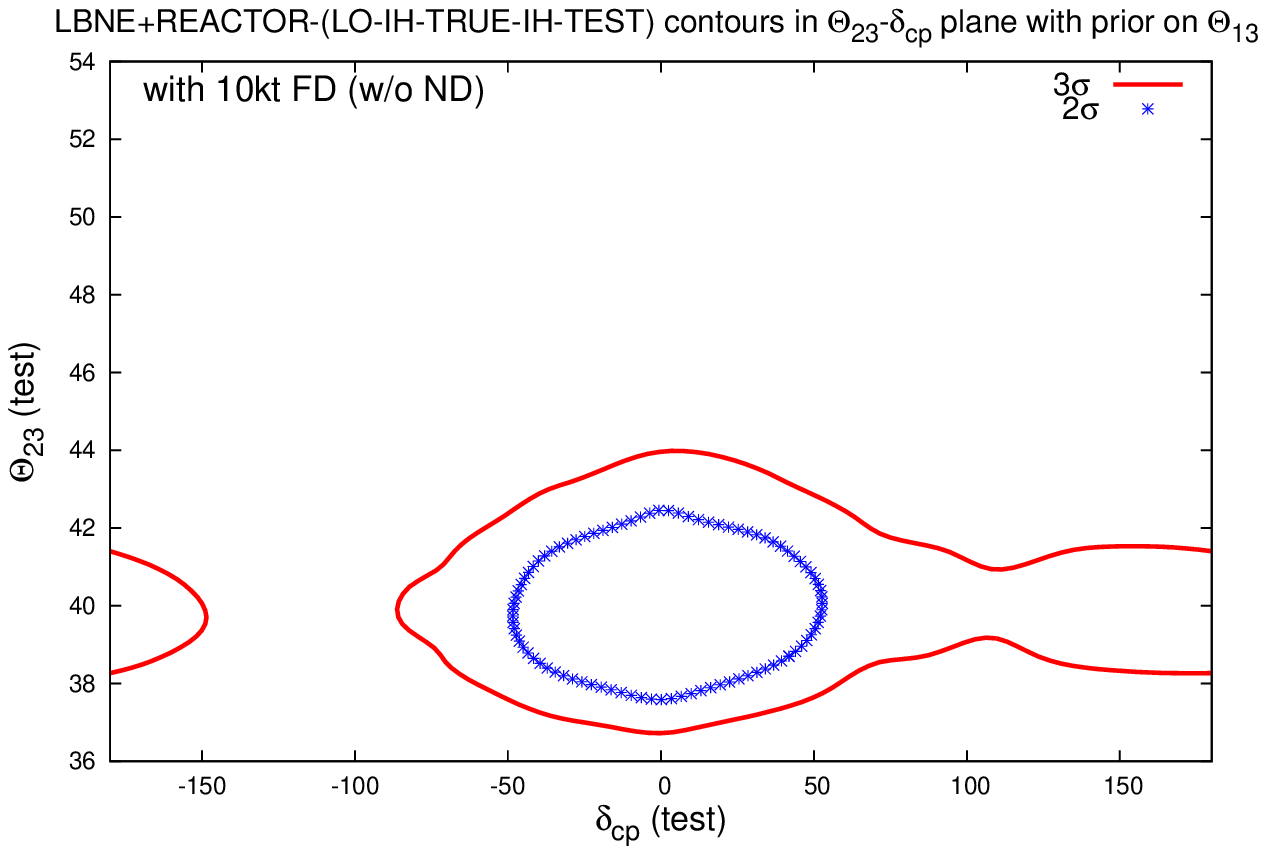}
\end{tabular}
\par\end{centering}

\caption{ Here in both the panels, we have shown our Reactors+LBNE (10 kt FD w/o ND) results in 
$\delta_{cp}(test)- \theta_{23}(test)$ plane in 2 d.o.f. for both NH (upper panel) and IH (lower panel). We have combined the 3 years of data from the three reactors with 10 years (5 years for $\nu$ and 5 years for $\bar{\nu}$) of date from LBNE. Prior is not added here.
}

\end{figure}

  In the upper panel of fig. 1, we have shown the marginalized octant sensitivity plots for the true values of $ \delta_{cp}=0.0$, both with normal and inverted hierarchy for LBNE with ND. In the lower panel of fig. 1, the octant sensitivity of 10 kt liquid argon detector without ND is studied and then compared with the reactor combined results. We have plotted the minimum value of $\chi^2$ as a function of true values of $\theta_{23}$. The Green plot represents the octant sensitivity of LBNE beamline. The thick dashed Green plot in upper panel is obtained by considering a prior of $\sigma_{\sin^22\theta_{13}} = 0.05 \times \sin^22\theta_{13} $.  Since the error in the measurement of $\theta_{13}$ is $\sigma_{\sin^22\theta_{13}} = 0.01$ as quoted by the recent reactor experiments, we can compare our combined plot with the thin Green LBNE plot.
It is observed that combining DB with LBNE improves the octant sensitivity noticeably. Effect of  DC+LBNE and RENO+LBNE is not impressive compared to DB+LBNE. There may be two reasons for this difference:

   1) Total thermal power output of DB is higher than RENO and DC. 
   
   2) DB employs four detector modules at its far site and two at the near site, each of which has a mass of 20 ton. Hence the total mass of DB FD is 80 ton and that of ND is 40 ton. RENO and DC has one far/near detector with a mass of 16.1/16.1 and 10/10 ton respectively. Therefore, more detector mass means more sensitivity in measurements, which in turn also improves octant sensitivity. 
  
In the lower panel of fig. 1, we have presented the results of octant sensitivity at LBNE without ND. It is observed that the octant sensitivity decreases when measured without ND. In fig. 2, we have shown the octant sensitivity of LBNE (for 10 kt FD with ND) with new global data. In fig. 3, we have compared the octant sensitivity of 5 years data from reactors with its 3 years data when combined with 10 years LBNE data. The difference between the plots is negligible, i.e. the octant sensitivity does not change noticeably as the data taking of reactor experiments increases from 3 to 5 years.
In fig. 4, we have presented our results of octant sensitivity for the 35 kt liquid argon detector with and without ND. Comparing fig. 1 and fig. 4, we have found that octant sensitivity improves if mass of the detector is increased.

We have also examined the possibility of resolving octant degeneracy present in LBNE beamline. The contours generated in $\delta_{cp}(test)-\theta_{23} (test)$ space for a true values of $\delta_{cp}=0$ and $\theta_{23}=$ $ 39.82$ $ (50.19)$ in LO (HO) clearly show the presence of both the octants (Fig. 5). While generating the contours, we have considered the test hierarchy to be the same as the true hierarchy and for the true values of $\theta_{23}$ in HO (LO) the test $\theta_{23}$ varies in the LO (HO).

In fig. 6, we have shown how the information on $\theta_{13}$ helps in resolving octant degeneracy. The figure in the left panel has been produced with LBNE data, and has two bands corresponding to $\theta_{23}$ and $(\pi/2)-\theta_{23}$, while the figure in the right, which is for LBNE with prior on $\theta_{13}$, has only one band corresponding to the true value of $\theta_{23}$. It means that, irrespective of the true value of $\delta_{cp}$, it is possible to determine the proper octant in the light of precise $\theta_{13}$ for a true value of $\theta_{23}$. In fig. 7, we have shown how the added prior affects the results. We have added prior on $\theta_{13}$ ( $\sigma_{\sin^22\theta_{13}}= 0.01$ ) with LBNE and as a result, the region corresponding to the test octant disappears at 2$\sigma$ cl and hence it is possible to pin-point the true octant.
 
 All the four plots in fig. 8 are generated for the combined data set i.e for DB+DC+RENO\\+LBNE data set. It is noticed that all the four plots are nearly similar to the respective four plots of fig. 7. So we can conclude that adding reactor experiments is equivalent to adding a proper prior on $\theta_{13}$ as far as octant sensitivity is concerned. In fig. 9 and fig. 10, we have shown our results for the octant degeneracy of LBNE10 (w/o ND) with and without prior on $\theta_{13}$. Adding prior on $\theta_{13}$ improves the results. In fig. 11, the effect of adding reactors to LBNE10 (w/o ND) is shown. It is found that adding reactors to LBNE10 can resolve octant degeneracy even without any ND. In the present scenario of LBNE, this result is really interesting.\\ As the area of the contours determines the allowed region and precision decreases with an increase in the allowed region, so by comparing fig. 9 with fig. 5, fig. 10 with fig. 7 and fig. 11 with fig. 8, we can conclude that the precision of the measurement increases when ND information is included in the calculations.

\section{Conclusion}

In this study, we have explored the possibility of resolving the octant ambiguity present in LBNE results by combining it with reactor data. The octant sensitivity of LBNE is found to increase significantly when combined with the reactor experiments. Out of the three reactor experiments Daya Bay, Double Chooz and RENO, the effect of adding DB information has the most significant effect, as it has massive detector facilities as well as high power output. Reactor experiments precisely measure the third mixing angle $\theta_{13}$. This information from reactor experiments in turn helps in determining the octant of $\theta_{23}$ precisely. We also note that adding reactors with LBNE is equivalent to adding a proper prior on $\theta_{13}$ with LBNE as in both ways one can resolve the octant degeneracy to the same extent. The comparative study of octant sensitivities of 10 kt and 35 kt detectors reveals that the sensitivity increases with an increase in detector mass. The presence of the Near Detector helps in reducing the systematic uncertainties and hence assists in improving the octant sensitivity. Also, the octant sensitivity changes negligibly with an increase in exposure of the reactor experiments. It may be noted that adding information from reactor experiments and ND also helps in improving CP violation sensitivity at LBNE \cite{51}.

\section{Acknowledgement}
We thank Raj Gandhi for extensive discussions, suggestions and for critically reading this manuscript. DD and KB thank the XI Plan Neutrino Project at HRI, Allahabad, for providing financial assistance to visit HRI, during which major parts of the work have been carried out. DD and KB thank Bipul Bhuyan, IIT, Guwahati, for providing his cluster for computation. PG thanks Srubabati Goswami, Sushant Raut and Monojit Ghosh for discussions regarding the octant degeneracy. DD also thanks Suprabh Prakash, Arnab Dasgupta, Animesh Chatterjee and Mehedi Masud for discussions regarding GLoBES.


\begin{thebibliography}{10}

\bibliographystyle{style}
\bibliography{bibfile}

\bibitem[1]{1} DAYA-BAY Collaboration, F. An et al., Phys.Rev.Lett. \textbf{108} (2012) 171803, [arXiv:1203.1669].

\bibitem[2]{2}RENO Collaboration, J. Ahn et al., Phys.Rev.Lett. \textbf{108} (2012) 191802, [arXiv:1204.0626].

\bibitem[3]{3} Double Chooz Collaboration, Y. Abe et al., [arXiv:1207.6632].

\bibitem[4]{4}Neutrino 2014, XXVI International Conference on Neutrino Physics and Astrophysics, June 2-7, 2014, Boston, U.S.A.
\bibitem[5]{5} RENO-50 Collaboration, R. collaboration, RENO-50, 
in International Workshop on RENO-50 toward Neutrino Mass Hierarchy, 2013.

\bibitem[6]{6} LBNE Collaboration, T. Akiri et al., [arXiv:1307.7335].

\bibitem[7]{7}T. Akiri et al. [LBNE Collaboration], arXiv:1110.6249 [hep-ex].

\bibitem[8]{8}T2K Collaboration, K. Abe et al., Phys.Rev.Lett. \textbf{107} (2011) 041801, [arXiv:1106.2822].

\bibitem[9]{9}MINOS Collaboration, P. Adamson et al., Phys.Rev.Lett. \textbf{107} (2011) 181802, [arXiv:1108.0015].

\bibitem[10]{10}D. Autiero, J. Aysto, A. Badertscher, L. B. Bezrukov, J. Bouchez, et al., JCAP \textbf{0711}, 011 (2007), 0705.0116.

\bibitem[11]{11}S. P. Mikheev and A. Y. Smirnov, Sov. J. Nucl. Phys. \textbf{42} (1985) 913–917.

\bibitem[12]{12}SNO Collaboration, B. Aharmim et al., arXiv:1109.0763.

\bibitem[13]{13}KamLAND Collaboration, S. Abe et al., Phys.Rev.Lett. \textbf{100} (2008) 221803, [arXiv:0801.4589].

\bibitem[14]{14}Super-Kamiokande Collaboration, R. Wendell et al., Phys.Rev. \textbf{D81} (2010) 092004, [arXiv:1002.3471].

\bibitem[15]{15}D. Forero, M. Tortola, and J. Valle, arXiv:1405.7540 [hep-ph].

\bibitem[16]{16}M. Freund, P. Huber, M. Lindner, Nucl. Phys \textbf{B615}, 331 (2001), hep-ph/0105071.
\bibitem[17]{17}P. Huber, M. Lindner, W. Winter, Nucl. Phys. \textbf{B645}, 3 (2002), hep-ph/0204352.
\bibitem[18]{18}K. Bora, D. Dutta, 2014 J. Phys.: Conf. Ser.(IOP, UK) \textbf{481} 012019 , arXiv:1209.1870.
\bibitem[19]{19}H. Minakata, H. Nunokawa, S.J Parke, Phys. Rev. \textbf{D66}, 093012 (2002), hep-ph/0208163.

\bibitem[20]{20}M. Koike, T. Ota, J. Sato, Phys. Rev \textbf{D65}, 053015 (2002), hep-ph/0011387.
\bibitem[21]{21} J. Burguet-Castell et. al, Nucl. Phys. \textbf{B608}, 301 (2001), hep-ph/0103258.

\bibitem[22]{22}H. Minakata and H. Nunokawa, JHEP \textbf{10}, 001 (2001), hep-ph/0108085.

\bibitem[23]{23}R. Gandhi, P. Ghoshal, S. Goswami, P. Mehta, and S. U. Sankar, Phys.Rev. \textbf{D73} (2006) 053001, [hep-ph/0411252].

\bibitem[24]{24}H. Minakata, M. Sonoyama, and H. Sugiyama, Phys.Rev. \textbf{D70} (2004)
113012, [hep-ph/0406073].
\bibitem[25]{25} NOvA Collaboration, arXiv:1209.0716 [hep-ex].
\bibitem[26]{26}S. K. Agarwalla, S. Prakash, and S. U. Sankar, JHEP \textbf{1307}, 131 (2013), arXiv:1301.2574.

\bibitem[27]{27}A. Chatterjee, P. Ghoshal, S. Goswami, and S. K. Raut, JHEP \textbf{1306}, 010 (2013), arXiv:1302.1370.

\bibitem[28]{28}K. Hiraide, H. Minakata, T. Nakaya, H. Nunokawa, H. Sugiyama, et al., Phys.Rev.\textbf{D73} (2006) 093008, [hep-ph/0601258].

\bibitem[29]{29}Takaaki Kajita, Hisakazu Minakata, Shoei Nakayama, and Hiroshi Nunokawa, Phys.Rev. \textbf{D75} (2007) 013006, [hep-ph/0609286]

\bibitem[30]{30}H. Minakata, H. Sugiyama, O. Yasuda, K. Inoue, F. Suekane; arXiv:hep-ph/0211111.

\bibitem[31]{31}Kendall B. M. Mahn, Michael H. Shaevitz;  Int.J.Mod.Phys.A21:3825-3844,2006; arXiv:hep-ex/0409028.

\bibitem[32]{32}Patrick Huber, Manfred Lindner, Thomas Schwetz, Walter Winter, JHEP 0911:044,2009, arXiv:0907.1896.

\bibitem[33]{33}M.C. Gonzalez-Garcia, M. Maltoni, A.Yu. Smirnov, Phys.Rev.\textbf{D70} (2004) 093005, arXiv:hep-ph/0408170.

\bibitem[34]{34} Sandhya Choubey, Probir Roy, Phys.Rev.\textbf{D73} (2006) 013006, arXiv:hep-ph/0509197.

\bibitem[35]{35}D. Indumathi, M.V.N. Murthy, G. Rajasekaran,Nita Sinha, Phys.Rev. \textbf{D74} (2006) 053004, arXiv:hep-ph/0603264.

\bibitem[36]{36} K. Abe,T. Abe el al, arXiv:1109.3262.
\bibitem[37]{37}Vernon Barger, Raj Gandhi, Pomita Ghoshal, Srubabati Goswami, Danny Marfatia, Suprabh Prakash, Sushant K. Raut, S. Uma Sankar, Phys.Rev.Lett.109:091801, 2012, arXiv:1203.6012.

 \bibitem[38]{38} E. K. Akhmedov, R. Johansson, M. Lindner, T. Ohlsson, and T. Schwetz, JHEP \textbf{0404} (2004) 078, [hep-ph/0402175].

\bibitem[39]{39}M. Freund,  Phys.Rev. \textbf{D64} (2001) 053003, [hep-ph/0103300].

\bibitem[40]{40}H. Minakata, Phys. Rev. \textbf{D52}, 6630 (1995) [arXiv:hep-ph/9503417].

\bibitem[41]{41}R. Gandhi, P. Ghoshal, S. Goswami, P. Mehta, S. U. Sankar, et al., Phys.Rev.\textbf{D76} (2007) 073012, [arXiv:0707.1723].

\bibitem[42]{42}J. K. Ahn, S. R. Baek, S. Choi, arXiv:hep-ph/1003.1391.

\bibitem[43]{43}K.Bora, D.Dutta, P. Ghoshal, JHEP \textbf{12} (2012) 025, arXiv:1206.2172.

\bibitem[44]{44}Vernon Barger, Atri Bhattacharya, Animesh Chatterjee, Raj Gandhi, Danny Marfatia, Mehedi Masud, Phys. Rev. \textbf{D89}, 011302 (2014), [arXiv:1307.2519].

\bibitem[45]{45}Elizabeth Worcester, talk presented at Neutrino Near Detector Workshop, July 28-28, Fermilab-2013. 

\bibitem[46]{46}  F. Ardellier et al., Double CHOOZ collaboration, hep-ex/0606025 .

\bibitem[47]{47}Xinheng Guo et al., Daya Bay proposal, arXiv:hep-ex/0701029.

\bibitem[48]{48}P. Huber, J. Kopp, M. Lindner, M. Rolinec, and W. Winter, Comput.Phys.Commun. \textbf{177} (2007) 432–438, [hep-ph/0701187].

\bibitem[49]{49}P. Huber, M. Lindner, and W. Winter,Phys.Commun. \textbf{167} (2005) 195, [hep-ph/0407333].

\bibitem[50]{50}D. Forero, M. Tortola, and J. Valle, Phys.Rev. \textbf{D86} (2012) 073012, [arXiv:1205.4018]

\bibitem[51]{51}Debajyoti Dutta, Kalpana Bora, [arXiv:1409.8248], accepted in MPLA.

\end{thebibliography}
\end{document}